\documentclass[prd,aps,nofootinbib,floatfix,10pt]{revtex4}

\usepackage{amsmath,graphicx,epsfig,amssymb,dsfont,mathtools}
\usepackage[usenames]{color}
\usepackage{ulem} 
\usepackage{bigstrut}
\usepackage{slashed}
\usepackage{multirow}
\usepackage{subfigure}

\allowdisplaybreaks

\begin{document}
	
	\title{Weak decays of doubly heavy baryons: the FCNC processes}
	
	\author{Zhi-Peng Xing$^{1}$, Zhen-Xing Zhao$^{1}$~\footnote{Email:star\_0027@sjtu.edu.cn}}
	
	\affiliation{$^{1}$ INPAC, Shanghai Key Laboratory for Particle Physics and Cosmology,\\
		MOE Key Laboratory for Particle Physics, Astrophysics and Cosmology,\\
		School of Physics and Astronomy, Shanghai Jiao-Tong University, Shanghai
		200240, P.R. China }
	\begin{abstract}
		The discovery of doubly heavy baryon provides us with a new platform
		for precisely testing Standard Model and searching for new physics.
		As a continuation of our previous works, we investigate the FCNC processes
		of doubly heavy baryons. Light-front approach is adopted
		to extract the form factors, in which the two spectator quarks are
		viewed as a diquark. Results for form factors are then used to predict
		some phenomenological observables, such as the decay width and the
		forward-backward asymmetry. We find that most of the branching ratios
		for $b\to s$ processes are $10^{-8}\sim10^{-7}$ and those for $b\to d$
		processes are $10^{-9}\sim10^{-8}$. The flavor SU(3) symmetry and
		symmetry breaking effects are explored. Parametric uncertainties are
		also investigated.
	\end{abstract}
	\maketitle
	
	\section{Introduction}
	
	Just one year ago, LHCb collaboration announced the discovery of a
	doubly charmed baryon $\Xi_{cc}^{++}$ with the mass~\cite{Aaij:2017ueg}
	\begin{equation}
	m_{\Xi_{cc}^{++}}=(3621.40\pm0.72\pm0.27\pm0.14){\rm MeV}.\label{eq:LHCb_measurement}
	\end{equation}
	Since then, great theoretical interests have been devoted to the study
	of doubly heavy baryons, some of them can be found in Refs.~\cite{Chen:2017sbg,Yu:2017zst,Wang:2017mqp,Li:2017cfz,Meng:2017udf,Wang:2017azm,Karliner:2017qjm,Gutsche:2017hux,Li:2017pxa,Guo:2017vcf,Lu:2017meb,Xiao:2017udy,Sharma:2017txj,Ma:2017nik,Meng:2017dni,Li:2017ndo,Wang:2017qvg,Shi:2017dto,Hu:2017dzi,1803.01476,Zhao:2018mrg}.
	Recently, some more new results were reported on $\Xi_{cc}^{++}$ by LHCb collaboration, including the first measurement of the lifetime~\cite{Aaij:2018wzf} and the first observation of the new decay mode $\Xi_{cc}^{++}\to \Xi_{c}^{+}\pi^{+}$~\cite{Aaij:2018gfl}. After discovering $\Xi_{cc}^{++}$ in the decay mode of $\Xi_{cc}^{++}\to\Lambda_{c}^{+}K^{-}\pi^{+}\pi^{+}$,
	LHCb collaboration is also continuing to search for the $\Xi_{cc}^{+}$
	and $\Xi_{bc}$ baryons~\cite{Traill:2017zbs}. Comprehensive theoretical
	studies on weak decays are highly demanded and our previous and forthcoming
	works aim to fill this gap. In our previous works~\cite{Wang:2017mqp,Zhao:2018mrg},
	we have presented the calculations of 1/2 to 1/2 and 1/2 to 3/2 weak
	decays. As a continuation, we investigate the flavor-changing neutral
	current (FCNC) processes in this work.
	
	FCNC processes are considered to be an ideal place to the precise
	test of Standard Model (SM) and the search for new physics (NP), while
	the discovery of the doubly heavy baryon provides us a new platform.
	$b\to d/s$ process in SM is induced by the loop effect, thus its decay
	width is small. NP effects manifest themselves in two different ways.
	One is to enhance the Wilson coefficients, and the other is to introduce
	new effective operators which are absent in the SM. The typical value
	of branching ratio for FCNC processes is $\sim10^{-6}$ for mesonic
	sector. However, the small branching ratio can be compensated by the
	high luminosity at the $B$ factories. Also, with the accumulation
	of data, we are in an increasingly better position to study these
	semi-leptonic process. Baryonic rare decay modes, which are also induced
	by $b\to d/s\,l^{+}l^{-}$ at the quark level, are also important as its
	mesonic counterparts. Serious attention is deserved, both theoretically
	and experimentally.
	
	A doubly heavy baryon is composed of two heavy quarks and one light
	quark. Light flavor SU(3) symmetry arranges them into the presentation
	$\boldsymbol{3}$. For $1/2^{+}$ doubly heavy baryons, we have $\Xi_{cc}^{++,+}$
	and $\Omega_{cc}^{+}$ in the $cc$ sector, $\Xi_{bb}^{0,-}$ and
	$\Omega_{bb}^{-}$ in the $bb$ sector, while there are two sets of
	baryons in the $bc$ sector depending on the symmetric property under
	the interchange of $b$ and $c$ quarks. For the symmetric case, the
	set is denoted by $\Xi_{bc}^{+,0}$and $\Omega_{bc}^{0}$, while for
	the asymmetric case, the set is denoted by $\Xi_{bc}^{\prime+,\prime0}$
	and $\Omega_{bc}^{\prime0}$.\footnote{The convention here for $bc$ sector is opposite to that in Ref. \cite{Brown:2014ena}.}
	In reality these two sets probably mix with each other, which will
	not be considered in this work.
	
\begin{figure}[!]
		\includegraphics[width=0.5\columnwidth]{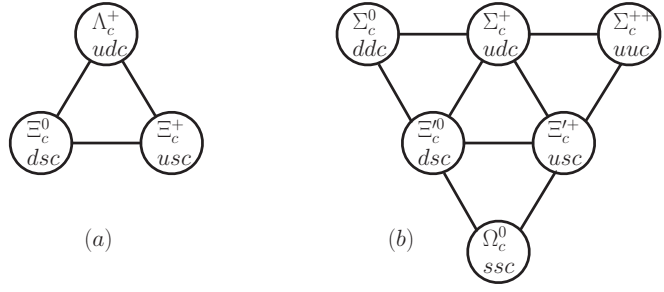} \caption{Spin-$1/2$ anti-triplets (panel a) and sextets (panel b) of charmed
			baryons. It is similar for the bottomed baryons.}
		\label{fig:singly_heavy}
	\end{figure}

	To be explicit, we will concentrate on the following FCNC decay modes
	of doubly heavy baryons. For $b\to s$ process,
	\begin{itemize}
		\item $bb$ sector
		\begin{align*}
		\Xi_{bb}^{0}(bbu) & \to\Xi_{b}^{0}(sbu)/\Xi_{b}^{\prime0}(sbu),\\
		\Xi_{bb}^{-}(bbd) & \to\Xi_{b}^{-}(sbd)/\Xi_{b}^{\prime-}(sbd),\\
		\Omega_{bb}^{-}(bbs) & \to\Omega_{b}^{-}(sbs),
		\end{align*}
		\item $bc$ sector
		\begin{align*}
		\Xi_{bc}^{+}(bcu)/\Xi_{bc}^{\prime+}(bcu) & \to\Xi_{c}^{+}(scu)/\Xi_{c}^{\prime+}(scu),\\
		\Xi_{bc}^{0}(bcd)/\Xi_{bc}^{\prime0}(bcd) & \to\Xi_{c}^{0}(scd)/\Xi_{c}^{\prime0}(scd),\\
		\Omega_{bc}^{0}(bcs)/\Omega_{bc}^{\prime0}(bcs) & \to\Omega_{c}^{0}(scs).
		\end{align*}
	\end{itemize}
For $b\to d$ process,
\begin{itemize}
\item $bb$ sector
\begin{align*}
\Xi_{bb}^{0}(bbu) & \to\Lambda_{b}^{0}(dbu)/\Sigma_{b}^{0}(dbu),\\
\Xi_{bb}^{-}(bbd) & \to\Sigma_{b}^{-}(dbd),\\
\Omega_{bb}^{-}(bbs) & \to\Xi_{b}^{-}(dbs)/\Xi_{b}^{\prime-}(dbs),
\end{align*}
\item $bc$ sector
\begin{align*}
\Xi_{bc}^{+}(bcu)/\Xi_{bc}^{\prime+}(bcu) & \to\Lambda_{c}^{+}(dcu)/\Sigma_{c}^{+}(dcu),\\
\Xi_{bc}^{0}(bcd)/\Xi_{bc}^{\prime0}(bcd) & \to\Sigma_{c}^{0}(dcd),\\
\Omega_{bc}^{0}(bcs)/\Omega_{bc}^{\prime0}(bcs) & \to\Xi_{c}^{0}(dcs)/\Xi_{c}^{\prime0}(dcs).
\end{align*}
\end{itemize}
	In the above, the quark components of the baryons have been explicitly
	presented in the brackets, and the quarks that participate in weak
	decay are put in the first place. Taking the $b\to s$ process in $bc$ sector as an example, the final baryons $\Xi_{c}^{+,0}$
	belong to the presentation of $\bar{\boldsymbol{3}}$, while $\Xi_{c}^{\prime+,\prime0}$
	and $\Omega_{c}^{0}$ belong to the presentation of $\boldsymbol{6}$,
	as can be seen from Fig.~\ref{fig:singly_heavy}.
	
\begin{figure}[!]
		\includegraphics[width=0.4\columnwidth]{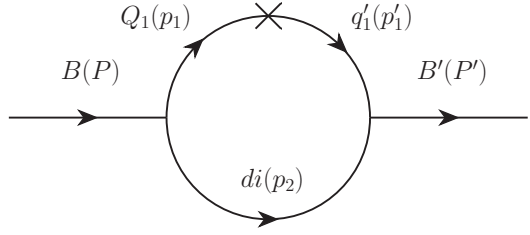} \caption{Feynman diagram for baryon-baryon transition in the diquark picture.
			$P^{(\prime)}$ is the momentum of the parent (daughter) baryon, $p_{1}^{(\prime)}$
			is the initial (final) quark momentum, $p_{2}$ is the diquark momentum
			and the cross mark denotes the weak interaction.}
		\label{fig:decay}
	\end{figure}

	Light front approach will be adopted to deal with the dynamics in
	the decay. This method has been widely used to study the mesonic decays
	~\cite{Jaus:1999zv,Jaus:1989au,Jaus:1991cy,Cheng:1996if,Cheng:2003sm,Cheng:2004yj,Ke:2009ed,Ke:2009mn,Cheng:2009ms,Lu:2007sg,Wang:2007sxa,Wang:2008xt,Wang:2008ci,Wang:2009mi,Chen:2009qk,Li:2010bb,Verma:2011yw,Shi:2016gqt}.
	Its application to baryonic sector can be found in Refs.~\cite{Ke:2007tg,Wei:2009np,Ke:2012wa,Zhu:2018jet,Ke:2017eqo}.
	As in our previous works, diquark picture is once again adopted, i.e.,
	the two spectator quarks are viewed as a whole system, as can be seen
	from Fig.~\ref{fig:decay}. The two spectator quarks form a scalar
	diquark or an axial-vector diquark. Generally speaking, both types
	of diquarks contribute to the decay process and their contribution
	weights can be determined by the wave functions of the baryons
	in the initial and final states.
	
	SU(3) analyses for FCNC processes will also be conducted. A quantitative
	predictions of SU(3) symmetry breaking effects will be performed within
	the light-front approach.
	
	The rest of the paper is arranged as follows. In Sec.~II, we will
	present the effective Hamiltonian responsible for the $b\to d/s\,l^{+}l^{-}$
	process. Then the framework of light-front approach under the diquark
	picture will be briefly introduced, then flavor-spin wave functions
	will also be discussed. Some phenomenological observables are collected
	in the last subsection of Sec.~II. Numerical results are shown
	in Sec.~III, including the results for form factors, decay widths,
	forward-backward asymmetry, the SU(3) symmetry breaking and the error estimates.
	A brief summary is given in the last section.
	
	\section{Theoretical framework}
	
	\subsection{The effective Hamiltonian}
	
	The effective Hamiltonian for $b\to sl^{+}l^{-}$ is given as
	\begin{equation}
	{\cal H}_{{\rm eff}}(b\to sl^{+}l^{-})=-\frac{G_{F}}{\sqrt{2}}V_{tb}V_{ts}^{*}\sum_{i=1}^{10}C_{i}(\mu)O_{i}(\mu).
	\end{equation}
	Here the explicit forms of the four-quark and the penguin operators
	$O_{i}$ can be found in Ref.~\cite{Buchalla:1995vs} and $C_{i}$
	are their corresponding Wilson coefficients, which are presented in Table~\ref{Tab:Wilson} in the leading logarithm approximation~\cite{Buchalla:1995vs}. The transition amplitude
	for ${\cal B}\to{\cal B}^{\prime}l^{+}l^{-}$ turns out to be
	\begin{eqnarray}
	{\cal M}({\cal B}\to{\cal B}^{\prime}l^{+}l^{-}) & = & -\frac{G_{F}}{\sqrt{2}}V_{tb}V_{ts}^{*}\frac{\alpha_{{\rm em}}}{2\pi}\Bigg\{\left(C_{9}^{{\rm eff}}(q^{2})\langle{\cal B}^{\prime}|\bar{s}\gamma_{\mu}(1-\gamma_{5})b|{\cal B}\rangle-2m_{b}C_{7}^{{\rm eff}}\langle{\cal B}^{\prime}|\bar{s}i\sigma_{\mu\nu}\frac{q^{\nu}}{q^{2}}(1+\gamma_{5})b|{\cal B}\rangle\right)\bar{l}\gamma^{\mu}l\nonumber \\
	&  & \qquad\qquad\qquad\quad+\,C_{10}\langle{\cal B}^{\prime}|\bar{s}\gamma_{\mu}(1-\gamma_{5})b|{\cal B}\rangle\bar{l}\gamma^{\mu}\gamma_{5}l\Bigg\}.\label{eq:the_amplitude}
	\end{eqnarray}
	Note that the sign before $C_{7}^{{\rm eff}}$ is different in literatures. Our result coincides with those in Refs.~\cite{Li:2009rc,Lu:2011jm}, but is different from that in Ref.~\cite{Giri:2005mt}. In Eq.~(\ref{eq:the_amplitude}), $C_{7}^{{\rm eff}}$ and $C_{9}^{{\rm eff}}$
	are defined by~\cite{Buras:1994dj}
	\begin{eqnarray}
	C_{7}^{{\rm eff}} & = & C_{7}-C_{5}/3-C_{6},\nonumber \\
	C_{9}^{{\rm eff}}(q^{2}) & = & C_{9}(\mu)+h(\hat{m}_{c},\hat{s})C_{0}-\frac{1}{2}h(1,\hat{s})(4C_{3}+4C_{4}+3C_{5}+C_{6})\nonumber \\
	&  & -\frac{1}{2}h(0,\hat{s})(C_{3}+3C_{4})+\frac{2}{9}(3C_{3}+C_{4}+3C_{5}+C_{6}),
	\end{eqnarray}
	with $\hat{s}=q^{2}/m_{b}^{2}$, $C_{0}=C_{1}+3C_{2}+3C_{3}+C_{4}+3C_{5}+C_{6}$,
	and $\hat{m}_{c}=m_{c}/m_{b}$. The auxiliary functions $h$ are given
	by
	\begin{eqnarray}
	h(z,\hat{s}) & = & -\frac{8}{9}\ln\frac{m_{b}}{\mu}-\frac{8}{9}\ln z+\frac{8}{27}+\frac{4}{9}x-\frac{2}{9}(2+x)|1-x|^{1/2}\times\begin{cases}
	\left(\ln\left|\frac{\sqrt{1-x}+1}{\sqrt{1-x}-1}\right|-i\pi\right), & x\equiv\frac{4z^{2}}{\hat{s}}<1\\
	2\arctan\frac{1}{\sqrt{x-1}}, & x\equiv\frac{4z^{2}}{\hat{s}}>1
	\end{cases},\nonumber \\
	h(0,\hat{s}) & = & -\frac{8}{9}\ln\frac{m_{b}}{\mu}-\frac{4}{9}\ln\hat{s}+\frac{8}{27}+\frac{4}{9}i\pi.
	\end{eqnarray}

	The effective Hamiltonian and transition amplitude for $b\to d$ process can be written down in a similar way.

	\begin{table}[!htbp]
		\caption{Wilson coefficients $C_{i}(m_{b})$ calculated in the leading logarithmic approximation, with $m_{W}=80.4$~GeV and $\mu=m_{b,{\rm pole}}$~\cite{Buchalla:1995vs}. }
		\label{Tab:Wilson} %
		\begin{tabular}{ccccccccc}
			\hline
			$C_{1}$  & $C_{2}$  & $C_{3}$  & $C_{4}$  & $C_{5}$  & $C_{6}$  & $C_{7}^{{\rm eff}}$  & $C_{9}$  & $C_{10}$\tabularnewline
			$1.107$  & $-0.248$  & $-0.011$  & $-0.026$  & $-0.007$  & $-0.031$  & $-0.313$  & $4.344$  & $-4.669$\tabularnewline
			\hline
		\end{tabular}
	\end{table}
	\subsection{Light-front approach}
	
	\label{subsec:light-front approach}
	
	Light-front approach for $1/2\to1/2$ FCNC transition will be briefly
	introduced in this subsection, including the definitions of the states
	for spin-1/2 baryons, and the extraction of form factors. More details
	can be found in Ref. \cite{Ke:2017eqo,Ke:2007tg}.
	
	In the light-front approach, the wave function of $1/2^{+}$ baryon
	with a scalar or an axial-vector diquark are expressed as
	\begin{eqnarray}
	|{\cal B}(P,S,S_{z})\rangle & = & \int\{d^{3}p_{1}\}\{d^{3}p_{2}\}2(2\pi)^{3}\delta^{3}(\tilde{P}-\tilde{p}_{1}-\tilde{p}_{2})\nonumber \\
	&  & \times\sum_{\lambda_{1},\lambda_{2}}\Psi^{SS_{z}}(\tilde{p}_{1},\tilde{p}_{2},\lambda_{1},\lambda_{2})|q_{1}(p_{1},\lambda_{1})(di)(p_{2},\lambda_{2})\rangle,\label{eq:state_vector}
	\end{eqnarray}
	where $q_{1}$ stands for $b/s$ quark in the initial/final state,
	and the diquark is denoted by $(di)$, which is composed of one $b$
	quark and one light quark. The momentum-space wave function $\Psi^{SS_{z}}$
	is given as
	\begin{equation}
	\Psi^{SS_{z}}(\tilde{p}_{1},\tilde{p}_{2},\lambda_{1},\lambda_{2})=\frac{A}{\sqrt{2(p_{1}\cdot\bar{P}+m_{1}M_{0})}}\bar{u}(p_{1},\lambda_{1})\Gamma u(\bar{P},S_{z})\phi(x,k_{\perp}),\label{eq:momentum_wave_function}
	\end{equation}
	with $A=1$ and $\Gamma=1$ for the case of a scalar diquark involved,
	and $A=\sqrt{\frac{3(m_{1}M_{0}+p_{1}\cdot\bar{P})}{3m_{1}M_{0}+p_{1}\cdot\bar{P}+2(p_{1}\cdot p_{2})(p_{2}\cdot\bar{P})/m_{2}^{2}}}$
	and $\Gamma=-\frac{1}{\sqrt{3}}\gamma_{5}\slashed\epsilon^{*}(p_{2},\lambda_{2})$
	for the case of an axial-vector diquark involved. A Gaussian-type
	function is usually adopted for $\phi$:
	\begin{equation}
	\phi=4\left(\frac{\pi}{\beta^{2}}\right)^{3/4}\sqrt{\frac{e_{1}e_{2}}{x_{1}x_{2}M_{0}}}\exp\left(\frac{-\vec{k}^{2}}{2\beta^{2}}\right).\label{eq:Gauss}
	\end{equation}
	
	Taking the $V-A$ current of $b\to s$ process as an example, the transition matrix element can be derived as
	\begin{eqnarray}
	&  & \langle{\cal B}^{\prime}(P^{\prime},S_{z}^{\prime})|\bar{s}\gamma_{\mu}(1-\gamma_{5})b|{\cal B}(P,S_{z})\rangle\nonumber \\
	& = & \int\{d^{3}p_{2}\}\frac{\varphi^{\prime}(x^{\prime},k_{\perp}^{\prime})\varphi(x,k_{\perp})}{2\sqrt{p_{1}^{+}p_{1}^{\prime+}(p_{1}\cdot\bar{P}+m_{1}M_{0})(p_{1}^{\prime}\cdot\bar{P}^{\prime}+m_{1}^{\prime}M_{0}^{\prime})}}\nonumber \\
	&  & \times\sum_{\lambda_{2}}\bar{u}(\bar{P}^{\prime},S_{z}^{\prime})\bar{\Gamma}^{\prime}(\slashed p_{1}^{\prime}+m_{1}^{\prime})\gamma_{\mu}(1-\gamma_{5})(\slashed p_{1}+m_{1})\Gamma u(\bar{P},S_{z}),\label{eq:matrix_element_1}
	\end{eqnarray}
	where
	\begin{equation}
	m_{1}=m_{b},\quad m_{1}^{\prime}=m_{s},\quad m_{2}=m_{(di)},
	\end{equation}
	and $\varphi^{(\prime)}=A^{(\prime)}\phi^{(\prime)}$, $p_{1}$ ($p_{1}^{\prime}$)
	denotes the four-momentum of the initial (final) quark, $P$ ($P^{\prime}$)
	stands for the four-momentum of ${\cal B}$ (${\cal B}^{\prime}$)
	in the initial (final) state. For the case of a scalar diquark involved,
	\begin{equation}
	\Gamma=\bar{\Gamma}^{\prime}=1,
	\end{equation}
	while for the case of an axial-vector diquark involved,
	\begin{align}
	\Gamma & =-\frac{1}{\sqrt{3}}\gamma_{5}\slashed\epsilon^{*}(p_{2},\lambda_{2})
	\end{align}
	and
	\begin{align}
	\bar{\Gamma}^{\prime} & =-\frac{1}{\sqrt{3}}\gamma_{5}\slashed\epsilon(p_{2},\lambda_{2}).
	\end{align}
	
	The transition matrix element $\langle{\cal B}^{\prime}(P^{\prime},S_{z}^{\prime})|\bar{s}\gamma_{\mu}(1-\gamma_{5})b|{\cal B}(P,S_{z})\rangle$
	can be parameterized as
	\begin{eqnarray}
	\langle{\cal B}^{\prime}(P^{\prime},S_{z}^{\prime})|\bar{s}\gamma_{\mu}b|{\cal B}(P,S_{z})\rangle & = & \bar{u}(P^{\prime},S_{z}^{\prime})\left[\gamma_{\mu}f_{1}(q^{2})+i\sigma_{\mu\nu}\frac{q^{\nu}}{M}f_{2}(q^{2})+\frac{q_{\mu}}{M}f_{3}(q^{2})\right]u(P,S_{z}),\nonumber \\
	\langle{\cal B}^{\prime}(P^{\prime},S_{z}^{\prime})|\bar{s}\gamma_{\mu}\gamma_{5}b|{\cal B}(P,S_{z})\rangle & = & \bar{u}(P^{\prime},S_{z}^{\prime})\left[\gamma_{\mu}g_{1}(q^{2})+i\sigma_{\mu\nu}\frac{q^{\nu}}{M}g_{2}(q^{2})+\frac{q_{\mu}}{M}g_{3}(q^{2})\right]\gamma_{5}u(P,S_{z}),\label{eq:matrix_element_2}
	\end{eqnarray}
	while $\langle{\cal B}^{\prime}(P^{\prime},S_{z}^{\prime})|\bar{s}i\sigma_{\mu\nu}q^{\nu}(1+\gamma_{5})b|{\cal B}(P,S_{z})\rangle$
	can be parameterized as
	\begin{eqnarray}
	\langle{\cal B}^{\prime}(P^{\prime},S_{z}^{\prime})|\bar{s}i\sigma_{\mu\nu}\frac{q^{\nu}}{M}b|{\cal B}(P,S_{z})\rangle & = & \bar{u}(P^{\prime},S_{z}^{\prime})\left[\gamma_{\mu}f_{1}^{T}(q^{2})+i\sigma_{\mu\nu}\frac{q^{\nu}}{M}f_{2}^{T}(q^{2})+\frac{q_{\mu}}{M}f_{3}^{T}(q^{2})\right]u(P,S_{z}),\nonumber \\
	\langle{\cal B}^{\prime}(P^{\prime},S_{z}^{\prime})|\bar{s}i\sigma_{\mu\nu}\frac{q^{\nu}}{M}\gamma_{5}b|{\cal B}(P,S_{z})\rangle & = & \bar{u}(P^{\prime},S_{z}^{\prime})\left[\gamma_{\mu}g_{1}^{T}(q^{2})+i\sigma_{\mu\nu}\frac{q^{\nu}}{M}g_{2}^{T}(q^{2})+\frac{q_{\mu}}{M}g_{3}^{T}(q^{2})\right]\gamma_{5}u(P,S_{z}).\label{eq:matrix_element_2p}
	\end{eqnarray}
	Here $q=P-P^{\prime}$, and $f_{i}^{(T)}$, $g_{i}^{(T)}$ are the
	form factors.
	
	It should be noted that $f_{1}^{T}$ and $f_{3}^{T}$ are not independent. Multiply the first equation of Eqs.~(\ref{eq:matrix_element_2p})
	by $q^{\mu}$ to yield
	\begin{equation}
	0=\bar{u}(P^{\prime},S_{z}^{\prime})\left[(M-M^{\prime})f_{1}^{T}+\frac{q^{2}}{M}f_{3}^{T}\right]u(P,S_{z}),
	\end{equation}
	and one obtains
	\begin{equation}
	f_{1}^{T}=-\frac{q^{2}}{M(M-M^{\prime})}f_{3}^{T}.\label{eq:f1T}
	\end{equation}
	In a similar way, one can obtain from the second equation of Eqs.~(\ref{eq:matrix_element_2p})
	\begin{equation}
	g_{1}^{T}=\frac{q^{2}}{M(M+M^{\prime})}g_{3}^{T}.\label{eq:g1T}
	\end{equation}
	
	Taking the extraction of $f_{i}$ as an example, these form factors can be extracted as follows \cite{Ke:2017eqo}. Multiplying the corresponding expressions in Eq.~(\ref{eq:matrix_element_1}) and Eqs.~(\ref{eq:matrix_element_2})
	by $\bar{u}(P,S_{z})(\Gamma^{\mu})_{i}u(P^{\prime},S_{z}^{\prime})$
	with $(\Gamma^{\mu})_{i}=\{\gamma^{\mu},P^{\mu},P^{\prime\mu}\}$
	respectively, and taking the approximation $P^{(\prime)}\to\bar{P}^{(\prime)}$
	within the integral, and then summing over the polarizations in the
	initial and final states, one can arrive at
	\begin{eqnarray}
	&  & {\rm Tr}\Bigg\{(\Gamma^{\mu})_{i}(\slashed P^{\prime}+M^{\prime})(\gamma_{\mu}f_{1}+i\sigma_{\mu\nu}\frac{q^{\nu}}{M}f_{2}+\frac{q_{\mu}}{M}f_{3})(\slashed P+M)\Bigg\}\nonumber \\
	& = & \int\{d^{3}p_{2}\}\frac{\varphi^{\prime}(x^{\prime},k_{\perp}^{\prime})\varphi(x,k_{\perp})}{2\sqrt{p_{1}^{+}p_{1}^{\prime+}(p_{1}\cdot\bar{P}+m_{1}M_{0})(p_{1}^{\prime}\cdot\bar{P}^{\prime}+m_{1}^{\prime}M_{0}^{\prime})}}\nonumber \\
	&  & \times\sum_{\lambda_{2}}{\rm Tr}\Bigg\{(\bar{\Gamma}^{\mu})_{i}(\bar{\slashed P}^{\prime}+M_{0}^{\prime})\bar{\Gamma}^{\prime}(\slashed p_{1}^{\prime}+m_{1}^{\prime})\gamma_{\mu}(\slashed p_{1}+m_{1})\Gamma(\bar{\slashed P}+M_{0})\Bigg\}
	\end{eqnarray}
	with $(\bar{\Gamma}^{\mu})_{i}=\{\gamma^{\mu},\bar{P}^{\mu},\bar{P}^{\prime\mu}\}$.
	Then $f_{i}$ can be determined by solving linear equations. The form
	factors $g_{i}$ can be obtained in a similar way. Only $f_{2,3}^{T}$
	or $g_{2,3}^{T}$ can be extracted in this way with $(\Gamma^{\mu})_{i}=\{\gamma^{\mu},P^{\mu}\}$,
	$f_{1}^{T}$ or $g_{1}^{T}$ is then obtained by Eq.~(\ref{eq:f1T}) or (\ref{eq:g1T}).
	
	\subsection{Flavor-spin wave functions}
	
	In fact, the flavor-spin wave functions are not
	taken into account in the last subsection. This problem will be fixed in this subsection.
	We consider first the initial states. For the doubly bottomed baryons,
	the wave functions are given as
	\begin{equation}
	\mathcal{B}_{bb}=\frac{1}{\sqrt{2}}\left[\left(-\frac{\sqrt{3}}{2}b^{1}(b^{2}q)_{S}+\frac{1}{2}b^{1}(b^{2}q)_{A}\right)+(b^{1}\leftrightarrow b^{2})\right],
	\end{equation}
	with $q=u$, $d$ or $s$ for $\Xi_{bb}^{0}$, $\Xi_{bb}^{-}$ or
	$\Omega_{bb}^{-}$, respectively. For the bottom-charm baryons, there
	are two sets of states, as discussed in Sec.~I. The wave functions of bottom-charm baryons with an axial-vector
	$bc$ diquark are given as
	\begin{align}
	\mathcal{B}_{bc} & =-\frac{\sqrt{3}}{2}b(cq)_{S}+\frac{1}{2}b(cq)_{A}\label{eq:flavor_spin_bc}
	\end{align}
	while those with a scalar $bc$ diquark are
	\begin{align}
	\mathcal{B}_{bc}^{\prime} & =-\frac{1}{2}b(cq)_{S}-\frac{\sqrt{3}}{2}b(cq)_{A}\label{eq:flavor_spin_bcp}
	\end{align}
	with $q=u$, $d$ or $s$ for $\Xi_{bc}^{(\prime)+}$, $\Xi_{bc}^{(\prime)0}$
	or $\Omega_{bc}^{(\prime)0}$, respectively.
	
	For the final states, the singly charmed baryon which belongs to anti-triplets are given as
\begin{align}
\Lambda_{c}^{+} & =-\frac{1}{2}d(cu)_{S}+\frac{\sqrt{3}}{2}d(cu)_{A},\nonumber \\
\Xi_{c}^{+} & =-\frac{1}{2}s(cu)_{S}+\frac{\sqrt{3}}{2}s(cu)_{A},\nonumber \\
\Xi_{c}^{0} & =-\frac{1}{2}s(cd)_{S}+\frac{\sqrt{3}}{2}s(cd)_{A}=\frac{1}{2}d(cs)_{S}-\frac{\sqrt{3}}{2}d(cs)_{A}.
\end{align}
For the sextet of singly charmed baryons, the following wave functions are needed
\begin{align}
\Sigma_{c}^{+} & =\frac{\sqrt{3}}{2}d(cu)_{S}+\frac{1}{2}d(cu)_{A},\nonumber \\
\Sigma_{c}^{0} & =\frac{1}{\sqrt{2}}\left[\frac{\sqrt{3}}{2}d^{1}(cd^{2})_{S}+\frac{1}{2}d^{1}(cd^{2})_{A}+(d^{1}\leftrightarrow d^{2})\right],\nonumber \\
\Xi_{c}^{\prime+} & =\frac{\sqrt{3}}{2}s(cu)_{S}+\frac{1}{2}s(cu)_{A},\nonumber \\
\Xi_{c}^{\prime0} & =\frac{\sqrt{3}}{2}s(cd)_{S}+\frac{1}{2}s(cd)_{A}=\frac{\sqrt{3}}{2}d(cs)_{S}+\frac{1}{2}d(cs)_{A},\nonumber \\
\Omega_{c}^{0} & =\frac{1}{\sqrt{2}}\left[\frac{\sqrt{3}}{2}s^{1}(cs^{2})_{S}+\frac{1}{2}s^{1}(cs^{2})_{A}+(s^{1}\leftrightarrow s^{2})\right].
\end{align}
	The final states of singly bottomed baryons can be written down in a similar way.

	Finally, the overlapping factors are determined by taking the inner
	product of the flavor-spin wave functions in the initial and final
	states. The corresponding results are collected in Table \ref{Tab:overlapping_factors} for both $b\to s$ and $b\to d$ processes.
	The physical form factors are then obtained by
	\begin{equation}
	F^{{\rm phy}}=c_{S}F_{S}+c_{A}F_{A},\label{eq:physical_ff}
	\end{equation}
	where $F_{S(A)}$ denotes the form factor $f_{i}$, $g_{i}$, $f_{i}^{T}$
	or $g_{i}^{T}$ with a scalar diquark (an axial-vector diquark) involved.
	
\begin{table}
\caption{Flavor-spin space overlapping factors for $b\to s$ and $b\to d$
processes. Taking the $\Xi_{bb}^{0}\to\Xi_{b}^{0}$ as an example, the physical
transition matrix elements can be evaluated as: $\langle\Xi_{b}^{0}|\Gamma_{\mu}|\Xi_{bb}^{0}\rangle=c_{S}\langle s[di]|\Gamma_{\mu}|b[di]\rangle+c_{A}\langle s\{di\}|\Gamma_{\mu}|b\{di\}\rangle$
with $c_{S}=\sqrt{6}/4$ and $c_{A}=\sqrt{6}/4$. Here $[di]$ and
$\{di\}$ denote a scalar and an axial-vector diquark, respectively.}
\label{Tab:overlapping_factors}
\begin{tabular}{c|c|c||c|c|c}
\hline
$b\to s$ process & $\langle s[di]|\Gamma_{\mu}|b[di]\rangle$  & $\langle s\{di\}|\Gamma_{\mu}|b\{di\}\rangle$ & $b\to d$ process & $\langle d[di]|\Gamma_{\mu}|b[di]\rangle$  & $\langle d\{di\}|\Gamma_{\mu}|b\{di\}\rangle$\tabularnewline
\hline
$\Xi_{bb}^{0}\to\Xi_{b}^{0}$  & \multirow{2}{*}{$\frac{\sqrt{6}}{4}$}  & \multirow{2}{*}{$\frac{\sqrt{6}}{4}$}  & $\Xi_{bb}^{0}\to\Lambda_{b}^{0}$  & $\frac{\sqrt{6}}{4}$  & $\frac{\sqrt{6}}{4}$\tabularnewline
\cline{4-6}
$\Xi_{bb}^{-}\to\Xi_{b}^{-}$  &   &   & $\Omega_{bb}^{-}\to\Xi_{b}^{-}$  & $-\frac{\sqrt{6}}{4}$  & $-\frac{\sqrt{6}}{4}$\tabularnewline
\hline
$\Xi_{bb}^{0}\to\Xi_{b}^{\prime0}$  &  \multirow{2}{*}{$-\frac{3\sqrt{2}}{4}$}  & \multirow{2}{*}{$\frac{\sqrt{2}}{4}$} & $\Xi_{bb}^{0}\to\Sigma_{b}^{0}$  & $-\frac{3\sqrt{2}}{4}$  & $\frac{\sqrt{2}}{4}$\tabularnewline
\cline{4-6}
$\Xi_{bb}^{-}\to\Xi_{b}^{\prime-}$  &  &  & $\Xi_{bb}^{-}\to\Sigma_{b}^{-}$  & $-\frac{3}{2}$  & $\frac{1}{2}$\tabularnewline
\hline
$\Omega_{bb}^{-}\to\Omega_{b}^{-}$  & $-\frac{3}{2}$  & $\frac{1}{2}$ & $\Omega_{bb}^{-}\to\Xi_{b}^{\prime-}$  & $-\frac{3\sqrt{2}}{4}$  & $\frac{\sqrt{2}}{4}$\tabularnewline
\hline
$\Xi_{bc}^{+}\to\Xi_{c}^{+}$  & \multirow{2}{*}{$\frac{\sqrt{3}}{4}$}  & \multirow{2}{*}{$\frac{\sqrt{3}}{4}$} & $\Xi_{bc}^{+}\to\Lambda_{c}^{+}$  & $\frac{\sqrt{3}}{4}$  & $\frac{\sqrt{3}}{4}$\tabularnewline
\cline{4-6}
$\Xi_{bc}^{0}\to\Xi_{c}^{0}$  &  &  & $\Omega_{bc}^{0}\to\Xi_{c}^{0}$  & $-\frac{\sqrt{3}}{4}$  & $-\frac{\sqrt{3}}{4}$\tabularnewline
\hline
$\Xi_{bc}^{+}\to\Xi_{c}^{\prime+}$  & \multirow{2}{*}{$-\frac{3}{4}$ } & \multirow{2}{*}{$\frac{1}{4}$ }& $\Xi_{bc}^{+}\to\Sigma_{c}^{+}$  & $-\frac{3}{4}$  & $\frac{1}{4}$\tabularnewline
\cline{4-6}
$\Xi_{bc}^{0}\to\Xi_{c}^{\prime0}$  &  &  & $\Xi_{bc}^{0}\to\Sigma_{c}^{0}$  & $-\frac{3\sqrt{2}}{4}$  & $\frac{\sqrt{2}}{4}$\tabularnewline
\hline
$\Omega_{bc}^{0}\to\Omega_{c}^{0}$  & $-\frac{3\sqrt{2}}{4}$  & $\frac{\sqrt{2}}{4}$ & $\Omega_{bc}^{0}\to\Xi_{c}^{\prime0}$  & $-\frac{3}{4}$  & $\frac{1}{4}$\tabularnewline
\hline
$\Xi_{bc}^{\prime+}\to\Xi_{c}^{+}$  & \multirow{2}{*}{$\frac{1}{4}$ } &\multirow{2}{*}{ $-\frac{3}{4} $ } & $\Xi_{bc}^{\prime+}\to\Lambda_{c}^{+}$  & $\frac{1}{4}$  & $-\frac{3}{4}$\tabularnewline
\cline{4-6}
$\Xi_{bc}^{\prime0}\to\Xi_{c}^{0}$  &  &  & $\Omega_{bc}^{\prime0}\to\Xi_{c}^{0}$  & $-\frac{1}{4}$  & $\frac{3}{4}$\tabularnewline
\hline
$\Xi_{bc}^{\prime+}\to\Xi_{c}^{\prime+}$  &\multirow{2}{*}{ $-\frac{\sqrt{3}}{4}$}  &\multirow{2}{*}{ $-\frac{\sqrt{3}}{4}$ }& $\Xi_{bc}^{\prime+}\to\Sigma_{c}^{+}$  & $-\frac{\sqrt{3}}{4}$  & $-\frac{\sqrt{3}}{4}$\tabularnewline
\cline{4-6}
$\Xi_{bc}^{\prime0}\to\Xi_{c}^{\prime0}$  &  &  & $\Xi_{bc}^{\prime0}\to\Sigma_{c}^{0}$  & $-\frac{\sqrt{6}}{4}$  & $-\frac{\sqrt{6}}{4}$\tabularnewline
\hline
$\Omega_{bc}^{\prime0}\to\Omega_{c}^{0}$  & $-\frac{\sqrt{6}}{4}$  & $-\frac{\sqrt{6}}{4}$ & $\Omega_{bc}^{\prime0}\to\Xi_{c}^{\prime0}$  & $-\frac{\sqrt{3}}{4}$  & $-\frac{\sqrt{3}}{4}$\tabularnewline
\hline
\end{tabular}
\end{table}
	
	\subsection{Phenomenological observables}
	
	The hadronic helicity amplitudes can be defined by
\begin{eqnarray}
H_{\lambda^{\prime},\lambda_{V}}^{V,\lambda} & \equiv & \Bigg(C_{9}^{{\rm eff}}(q^{2})\langle{\cal B}^{\prime}|\bar{s}\gamma^{\mu}(1-\gamma_{5})b|{\cal B}\rangle-C_{7}^{{\rm eff}}2m_{b}\langle{\cal B}^{\prime}|\bar{s}i\sigma^{\mu\nu}\frac{q_{\nu}}{q^{2}}(1+\gamma_{5})b|{\cal B}\rangle\Bigg)\epsilon_{\mu}^{*}(\lambda_{V}),\nonumber \\
H_{\lambda^{\prime},t}^{V,\lambda} & \equiv & \Bigg(C_{9}^{{\rm eff}}(q^{2})\langle{\cal B}^{\prime}|\bar{s}\gamma^{\mu}(1-\gamma_{5})b|{\cal B}\rangle\Bigg)\frac{q_{\mu}}{\sqrt{q^{2}}},
\label{eq:HV}
\end{eqnarray}
	and
\begin{eqnarray}
H_{\lambda^{\prime},\lambda_{V}}^{A,\lambda} & \equiv & \Bigg(C_{10}\langle{\cal B}^{\prime}|\bar{s}\gamma^{\mu}(1-\gamma_{5})b|{\cal B}\rangle\Bigg)\epsilon_{\mu}^{*}(\lambda_{V}),\nonumber \\
H_{\lambda^{\prime},t}^{A,\lambda} & \equiv & \Bigg(C_{10}\langle{\cal B}^{\prime}|\bar{s}\gamma^{\mu}(1-\gamma_{5})b|{\cal B}\rangle\Bigg)\frac{q_{\mu}}{\sqrt{q^{2}}},
\end{eqnarray}
where $\epsilon_{\mu}$ ($q_{\mu}$) is the polarization vector (four-momentum) for an intermediate vector particle, $\lambda_{V}$ denotes its polarization, $\lambda^{(\prime)}$ is the polarization of the baryon in the initial (final) state.
	Hereafter the superscript ``$V$'' (``$A$'') always means that its
	corresponding leptonic counterpart is $\bar{l}\gamma^{\mu}l$ ($\bar{l}\gamma^{\mu}\gamma_{5}l$).
	It should not be confused with the notation of the vector current (axial-vector current) in the hadronic matrix element.
	
	Note that Eqs.~(\ref{eq:matrix_element_2}) and (\ref{eq:matrix_element_2p})
	have the same parameterization, so it is convenient to introduce the
	following notations:
	\begin{eqnarray}
	F_{i}^{V}(q^{2}) & \equiv & C_{9}^{{\rm eff}}(q^{2})f_{i}(q^{2})-C_{7}^{{\rm eff}}\frac{2m_{b}M}{q^{2}}f_{i}^{T}(q^{2}),\nonumber \\
	G_{i}^{V}(q^{2}) & \equiv & C_{9}^{{\rm eff}}(q^{2})g_{i}(q^{2})+C_{7}^{{\rm eff}}\frac{2m_{b}M}{q^{2}}g_{i}^{T}(q^{2})
	\end{eqnarray}
	and
	\begin{eqnarray}
	F_{i}^{A}(q^{2}) & \equiv & C_{10}f_{i}(q^{2}),\nonumber \\
	G_{i}^{A}(q^{2}) & \equiv & C_{10}g_{i}(q^{2}).
	\end{eqnarray}
	Then the $\Gamma^{\mu}$ and $\Gamma^{\mu}\gamma_{5}$ parts in Eq.~(\ref{eq:HV}) are calculated respectively as:
	\begin{align}
	HV_{\frac{1}{2},0}^{V,-\frac{1}{2}} & =-i\frac{\sqrt{Q_{-}}}{\sqrt{q^{2}}}\left((M+M^{\prime})F_{1}^{V}-\frac{q^{2}}{M}F_{2}^{V}\right),\nonumber \\
	HV_{\frac{1}{2},1}^{V,\frac{1}{2}} & =i\sqrt{2Q_{-}}\left(-F_{1}^{V}+\frac{M+M^{\prime}}{M}F_{2}^{V}\right),\nonumber \\
	HA_{\frac{1}{2},0}^{V,-\frac{1}{2}} & =-i\frac{\sqrt{Q_{+}}}{\sqrt{q^{2}}}\left((M-M^{\prime})G_{1}^{V}+\frac{q^{2}}{M}G_{2}^{V}\right),\nonumber \\
	HA_{\frac{1}{2},1}^{V,\frac{1}{2}} & =i\sqrt{2Q_{+}}\left(-G_{1}^{V}-\frac{M-M^{\prime}}{M}G_{2}^{V}\right)
	\end{align}
	and
	\begin{eqnarray}
	HV_{-\lambda^{\prime},-\lambda_{V}}^{V,-\lambda} & = & HV_{\lambda^{\prime},\lambda_{V}}^{V,\lambda},\nonumber \\
	HA_{-\lambda^{\prime},-\lambda_{V}}^{V,-\lambda} & = & -HA_{\lambda^{\prime},\lambda_{V}}^{V,\lambda}.
	\end{eqnarray}
	The total hadronic helicity amplitude is then given by
	\begin{equation}
	H_{\lambda^{\prime},\lambda_{V}}^{V,\lambda}=HV_{\lambda^{\prime},\lambda_{V}}^{V,\lambda}-HA_{\lambda^{\prime},\lambda_{V}}^{V,\lambda}.
	\end{equation}
	$H_{\lambda^{\prime},\lambda_{V}}^{A,\lambda}$ has the complete the
	same form as the corresponding $H_{\lambda^{\prime},\lambda_{V}}^{V,\lambda}$
	but with the following replacements:
	\begin{eqnarray}
	F_{i}^{V} & \to & F_{i}^{A},\nonumber \\
	G_{i}^{V} & \to & G_{i}^{A}.
	\end{eqnarray}
	In addition, the timelike polarizations for $H^{A}$ are also needed
	\begin{align}
	HV_{-\frac{1}{2},t}^{A,\frac{1}{2}}=HV_{\frac{1}{2},t}^{A,-\frac{1}{2}}=-i\frac{\sqrt{Q_{+}}}{\sqrt{q^{2}}}\left((M-M^{\prime})F_{1}^{A}+\frac{q^{2}}{M}F_{3}^{A}\right),\nonumber \\
	-HA_{-\frac{1}{2},t}^{A,\frac{1}{2}}=HA_{\frac{1}{2},t}^{A,-\frac{1}{2}}=-i\frac{\sqrt{Q_{-}}}{\sqrt{q^{2}}}\left((M+M^{\prime})G_{1}^{A}-\frac{q^{2}}{M}G_{3}^{A}\right)
	\end{align}
	and
	\begin{equation}
	H_{\lambda^{\prime},t}^{A,\lambda}=HV_{\lambda^{\prime},t}^{A,\lambda}-HA_{\lambda^{\prime},t}^{A,\lambda}.
	\end{equation}
	
	Finally, the angular distribution is given by the following expression
	\begin{eqnarray}
	\frac{d^{2}\Gamma}{dq^{2}d\cos\theta} & = & \frac{|\vec{P}^{\prime}||\vec{p}_{1}|}{16(2\pi)^{3}M^{2}\sqrt{q^{2}}}\overline{|{\cal M}|^{2}}.\label{eq:angular_distribution}
	\end{eqnarray}
	Here the squared amplitude is
	\begin{equation}
	\overline{|{\cal M}|^{2}}=\frac{1}{2}|\lambda|^{2}(I_{0}+I_{1}\cos\theta+I_{2}\cos2\theta)\label{eq:sq_M_bar}
	\end{equation}
	with
	\begin{equation}
	\lambda\equiv\frac{G_{F}}{\sqrt{2}}V_{tb}V_{ts}^{*}\frac{\alpha_{{\rm em}}}{2\pi}
	\end{equation}
	and
	\begin{eqnarray}
	I_{0} & = & (q^{2}+4m_{l}^{2})(|H_{-\frac{1}{2},0}^{V,\frac{1}{2}}|^{2}+|H_{\frac{1}{2},0}^{V,-\frac{1}{2}}|^{2})+(\frac{3}{2}q^{2}+2m_{l}^{2})(|H_{\frac{1}{2},1}^{V,\frac{1}{2}}|^{2}+|H_{-\frac{1}{2},-1}^{V,-\frac{1}{2}}|^{2})\nonumber \\
	&  & +(q^{2}-4m_{l}^{2})(\frac{3}{2}|H_{\frac{1}{2},1}^{A,\frac{1}{2}}|^{2}+\frac{3}{2}|H_{-\frac{1}{2},-1}^{A,-\frac{1}{2}}|^{2}+|H_{-\frac{1}{2},0}^{A,\frac{1}{2}}|^{2}+|H_{\frac{1}{2},0}^{A,-\frac{1}{2}}|^{2})\nonumber \\
	&  & +8m_{l}^{2}(|H_{-\frac{1}{2},t}^{A,\frac{1}{2}}|^{2}+|H_{\frac{1}{2},t}^{A,-\frac{1}{2}}|^{2}),\nonumber \\
	I_{1} & = & 4\sqrt{q^{2}(q^{2}-4m_{l}^{2})}{\rm Re}(H_{\frac{1}{2},1}^{A,\frac{1}{2}*}H_{\frac{1}{2},1}^{V,\frac{1}{2}}-H_{-\frac{1}{2},-1}^{A,-\frac{1}{2}*}H_{-\frac{1}{2},-1}^{V,-\frac{1}{2}}),\nonumber \\
	I_{2} & = & \frac{1}{2}(q^{2}-4m_{l}^{2})(|H_{\frac{1}{2},1}^{V,\frac{1}{2}}|^{2}+|H_{-\frac{1}{2},-1}^{V,-\frac{1}{2}}|^{2}-2|H_{-\frac{1}{2},0}^{V,\frac{1}{2}}|^{2}-2|H_{\frac{1}{2},0}^{V,-\frac{1}{2}}|^{2}\nonumber \\
	&  & \qquad\qquad\quad+|H_{\frac{1}{2},1}^{A,\frac{1}{2}}|^{2}+|H_{-\frac{1}{2},-1}^{A,-\frac{1}{2}}|^{2}-2|H_{-\frac{1}{2},0}^{A,\frac{1}{2}}|^{2}-2|H_{\frac{1}{2},0}^{A,-\frac{1}{2}}|^{2}).
	\end{eqnarray}
	
	The differential decay width is given as
	\begin{eqnarray}
	\frac{d\Gamma}{dq^{2}} & = & \frac{d\Gamma_{L}}{dq^{2}}+\frac{d\Gamma_{T}}{dq^{2}},
	\end{eqnarray}
	where the $q^{2}$ is the invariant mass of the dilepton and the longitudinally
	and transversely polarized decay widths are respectively
	\begin{eqnarray}
	\frac{d\Gamma_{L}}{dq^{2}} & = & |\lambda|^{2}\frac{|\vec{P}^{\prime}||\vec{p}_{1}|}{12(2\pi)^{3}M^{2}\sqrt{q^{2}}}\Big\{(q^{2}+2m_{l}^{2})(|H_{-\frac{1}{2},0}^{V,\frac{1}{2}}|^{2}+|H_{\frac{1}{2},0}^{V,-\frac{1}{2}}|^{2})\nonumber \\
	&  & \qquad\qquad\qquad\qquad+(q^{2}-4m_{l}^{2})(|H_{-\frac{1}{2},0}^{A,\frac{1}{2}}|^{2}+|H_{\frac{1}{2},0}^{A,-\frac{1}{2}}|^{2})\nonumber \\
	&  & \qquad\qquad\qquad\qquad+6m_{l}^{2}(|H_{-\frac{1}{2},t}^{A,\frac{1}{2}}|^{2}+|H_{\frac{1}{2},t}^{A,-\frac{1}{2}}|^{2})\Big\},\\
	\frac{d\Gamma_{T}}{dq^{2}} & = & |\lambda|^{2}\frac{|\vec{P}^{\prime}||\vec{p}_{1}|}{12(2\pi)^{3}M^{2}\sqrt{q^{2}}}\Big\{(q^{2}+2m_{l}^{2})(|H_{\frac{1}{2},1}^{V,\frac{1}{2}}|^{2}+|H_{-\frac{1}{2},-1}^{V,-\frac{1}{2}}|^{2})\nonumber \\
	&  & \qquad\qquad\qquad\qquad+(q^{2}-4m_{l}^{2})(|H_{\frac{1}{2},1}^{A,\frac{1}{2}}|^{2}+|H_{-\frac{1}{2},-1}^{A,-\frac{1}{2}}|^{2})\Big\}.
	\end{eqnarray}
	
	The normalized differential forward-backward asymmetry is defined
	by
	\begin{equation}
	\frac{d\bar{A}_{FB}}{dq^{2}}\equiv\frac{(\int_{0}^{1}-\int_{-1}^{0})d\cos\theta\frac{d^{2}\Gamma}{dq^{2}d\cos\theta}}{(\int_{0}^{1}+\int_{-1}^{0})d\cos\theta\frac{d^{2}\Gamma}{dq^{2}d\cos\theta}}.\label{eq:def_AFB}
	\end{equation}
	Then one can obtain
	\begin{equation}
	\frac{d\bar{A}_{FB}}{dq^{2}}=\frac{I_{1}}{2(I_{0}-I_{2}/3)}
	\end{equation}
	when substituting Eqs.~(\ref{eq:angular_distribution}) and (\ref{eq:sq_M_bar}) into Eq.~(\ref{eq:def_AFB}).
	
	\section{Numerical results and discussions}
	
	\subsection{Inputs}
	
	The constituent quark masses are given as (in units of GeV)~\cite{Lu:2007sg,Wang:2007sxa,Wang:2008xt,Wang:2008ci,Wang:2009mi,Chen:2009qk,Li:2010bb,Verma:2011yw,Shi:2016gqt}
	\begin{equation}
	m_{u}=m_{d}=0.25,\quad m_{s}=0.37,\quad m_{c}=1.4,\quad m_{b}=4.8.
	\end{equation}
	The masses of the scalar and axial-vector diquarks are approximated
	by $m_{[Qq]}=m_{\{Qq\}}=m_{Q}+m_{q}$. The shape parameters $\beta$
	in Eq.~(\ref{eq:Gauss}) are given as (in units of GeV) \cite{Cheng:2003sm}
\begin{eqnarray}
 &  & \beta_{d[cq]}=\beta_{d\{cq\}}=0.470,\quad\beta_{s[cq]}=\beta_{s\{cq\}}=0.535,\quad\beta_{b[cq]}=\beta_{b\{cq\}}=0.886,\nonumber \\
 &  & \beta_{d[bq]}=\beta_{d\{bq\}}=0.562,\quad\beta_{s[bq]}=\beta_{s\{bq\}}=0.623,\quad\beta_{b[bq]}=\beta_{b\{bq\}}=1.472,
\end{eqnarray}
	where $q=u,d,s$.
	
	The masses and lifetimes of the parent baryons are collected in Table
	\ref{Tab:parent_baryons}~\cite{Brown:2014ena,Karliner:2014gca,Kiselev:2001fw}.
	The masses of the daughter baryons are given in Table \ref{Tab:daughter_baryons}~\cite{Olive:2016xmw}.
	Fermi constant and CKM matrix elements are give as~\cite{Olive:2016xmw}
	\begin{align}
	& G_{F}=1.166\times10^{-5}\ {\rm GeV}^{-2},\nonumber \\
	& |V_{tb}|=0.999,\quad|V_{ts}|=0.0403,\quad|V_{td}|=0.00875.\label{eq:GFCKM}
	\end{align}
	
	\begin{table}[!htb]
		\caption{Masses (in units of GeV) and lifetimes (in units of fs) of doubly
			heavy baryons. We have quoted the results from Refs.~\cite{Brown:2014ena,Karliner:2014gca,Kiselev:2001fw}.}
		\label{Tab:parent_baryons} %
		\begin{tabular}{c|c|c|c|c|c|c}
			\hline
			baryons  & $\Xi_{bc}^{(\prime)+}$  & $\Xi_{bc}^{(\prime)0}$  & $\Omega_{bc}^{(\prime)0}$  & $\Xi_{bb}^{0}$  & $\Xi_{bb}^{-}$  & $\Omega_{bb}^{-}$ \tabularnewline
			\hline
			masses  & $6.943$ \cite{Brown:2014ena}  & $6.943$ \cite{Brown:2014ena}  & $6.998$ \cite{Brown:2014ena}  & $10.143$\cite{Brown:2014ena}  & $10.143$ \cite{Brown:2014ena}  & $10.273$\cite{Brown:2014ena}\tabularnewline
			\hline
			lifetimes  & $244$ \cite{Karliner:2014gca}  & $93$ \cite{Karliner:2014gca}  & $220$ \cite{Kiselev:2001fw}  & $370$ \cite{Karliner:2014gca}  & $370$ \cite{Karliner:2014gca}  & $800$\cite{Kiselev:2001fw}\tabularnewline
			\hline
		\end{tabular}
	\end{table}
	
\begin{table}[!]
\caption{Masses (in units of GeV) of baryons in the final states~\cite{Olive:2016xmw}. }
\label{Tab:daughter_baryons} %
\begin{tabular}{c|c|c|c|c|c|c|c}
\hline
$\Lambda_{c}^{+}$  & $\Xi_{c}^{+}$  & $\Xi_{c}^{0}$  & $\Sigma_{c}^{+}$  & $\Sigma_{c}^{0}$  & $\Xi_{c}^{\prime+}$  & $\Xi_{c}^{\prime0}$  & $\Omega_{c}^{0}$ \tabularnewline
\hline
$2.286$  & $2.468$  & $2.471$  & $2.453$  & $2.454$  & $2.576$  & $2.578$  & $2.695$ \tabularnewline
\hline
$\Lambda_{b}^{0}$  & $\Xi_{b}^{0}$  & $\Xi_{b}^{-}$  & $\Sigma_{b}^{0}$  & $\Sigma_{b}^{-}$  & $\Xi_{b}^{\prime0}$  & $\Xi_{b}^{\prime-}$  & $\Omega_{b}^{-}$ \tabularnewline
\hline
$5.620$  & $5.793$  & $5.795$  & $5.814$  & $5.816$  & $5.935$  & $5.935$  & $6.046$ \tabularnewline
\hline
\end{tabular}
\end{table}

	\subsection{Results for form factors}
	
	To access the $q^{2}$-distribution, the following single pole structure
	is assumed for form factors
	\begin{equation}
	F(q^{2})=\frac{F(0)}{1-\frac{q^{2}}{m_{{\rm pole}^{2}}}}.\label{eq:single_pole}
	\end{equation}
	Here $F(0)$ is the value of the form factor at $q^{2}=0$, and the numerical results for $f^{(T)}_{i}$ and $g^{(T)}_{i}$ predicted by the light-front approach are collected
	in Tables \ref{Tab:fg_bb} to \ref{Tab:fTgT_bc} for $b\to s$ process and Tables \ref{Tab:fg_bb_b2d} to \ref{Tab:fTgT_bc_b2d} for $b\to d$ process. $m_{{\rm pole}}$
	is taken as $5.37$ GeV for $b\to s$ process and $5.28$ GeV for $b\to d$ process, which, in practice, are taken as the masses
	of $B_{s}$ and $B$ mesons, respectively. The discussion for the validity of this assumption
	can be found in our previous work~\cite{Shi:2016gqt}.
	
	The physical form factors can then be obtained by Eq.~(\ref{eq:physical_ff})
	and Eq.~(\ref{eq:single_pole}).
	
	\begin{table}
		\caption{Values of form factors $f_{i}$ and $g_{i}$ at $q^{2}=0$ for $b\to s$
			process in $bb$ sector. The left (right) half of the table corresponds
			to a scalar diquark (an axial-vector diquark) involved case.}
		\label{Tab:fg_bb} %
		\begin{tabular}{c|r|c|r||c|r|c|r}
			\hline
			$F$  & $F(0)$  & $F$  & $F(0)$  & $F$  & $F(0)$  & $F$  & $F(0)$ \tabularnewline
			\hline
			$f_{1,S}^{\Xi_{bb}\to\Xi_{b}}$  & $0.141$  & $g_{1,S}^{\Xi_{bb}\to\Xi_{b}}$  & $0.122$  & $f_{1,A}^{\Xi_{bb}\to\Xi_{b}}$  & $0.138$  & $g_{1,A}^{\Xi_{bb}\to\Xi_{b}}$  & $-0.030$ \tabularnewline
			$f_{2,S}^{\Xi_{bb}\to\Xi_{b}}$  & $-0.189$  & $g_{2,S}^{\Xi_{bb}\to\Xi_{b}}$  & $0.056$  & $f_{2,A}^{\Xi_{bb}\to\Xi_{b}}$  & $0.132$  & $g_{2,A}^{\Xi_{bb}\to\Xi_{b}}$  & $-0.055$ \tabularnewline
			$f_{3,S}^{\Xi_{bb}\to\Xi_{b}}$  & $0.016$  & $g_{3,S}^{\Xi_{bb}\to\Xi_{b}}$  & $-0.406$  & $f_{3,A}^{\Xi_{bb}\to\Xi_{b}}$  & $-0.068$  & $g_{3,A}^{\Xi_{bb}\to\Xi_{b}}$  & $0.261$ \tabularnewline
			\hline
			$f_{1,S}^{\Xi_{bb}\to\Xi_{b}^{\prime}}$  & $0.143$  & $g_{1,S}^{\Xi_{bb}\to\Xi_{b}^{\prime}}$  & $0.130$  & $f_{1,A}^{\Xi_{bb}\to\Xi_{b}^{\prime}}$  & $0.140$  & $g_{1,A}^{\Xi_{bb}\to\Xi_{b}^{\prime}}$  & $-0.031$ \tabularnewline
			$f_{2,S}^{\Xi_{bb}\to\Xi_{b}^{\prime}}$  & $-0.202$  & $g_{2,S}^{\Xi_{bb}\to\Xi_{b}^{\prime}}$  & $0.024$  & $f_{2,A}^{\Xi_{bb}\to\Xi_{b}^{\prime}}$  & $0.138$  & $g_{2,A}^{\Xi_{bb}\to\Xi_{b}^{\prime}}$  & $-0.048$ \tabularnewline
			$f_{3,S}^{\Xi_{bb}\to\Xi_{b}^{\prime}}$  & $0.003$  & $g_{3,S}^{\Xi_{bb}\to\Xi_{b}^{\prime}}$  & $-0.316$  & $f_{3,A}^{\Xi_{bb}\to\Xi_{b}^{\prime}}$  & $-0.082$  & $g_{3,A}^{\Xi_{bb}\to\Xi_{b}^{\prime}}$  & $0.249$ \tabularnewline
			\hline
			$f_{1,S}^{\Omega_{bb}^{-}\to\Omega_{b}^{-}}$  & $0.139$  & $g_{1,S}^{\Omega_{bb}^{-}\to\Omega_{b}^{-}}$  & $0.125$  & $f_{1,A}^{\Omega_{bb}^{-}\to\Omega_{b}^{-}}$  & $0.136$  & $g_{1,A}^{\Omega_{bb}^{-}\to\Omega_{b}^{-}}$  & $-0.030$ \tabularnewline
			$f_{2,S}^{\Omega_{bb}^{-}\to\Omega_{b}^{-}}$  & $-0.198$  & $g_{2,S}^{\Omega_{bb}^{-}\to\Omega_{b}^{-}}$  & $0.028$  & $f_{2,A}^{\Omega_{bb}^{-}\to\Omega_{b}^{-}}$  & $0.134$  & $g_{2,A}^{\Omega_{bb}^{-}\to\Omega_{b}^{-}}$  & $-0.048$ \tabularnewline
			$f_{3,S}^{\Omega_{bb}^{-}\to\Omega_{b}^{-}}$  & $0.003$  & $g_{3,S}^{\Omega_{bb}^{-}\to\Omega_{b}^{-}}$  & $-0.332$  & $f_{3,A}^{\Omega_{bb}^{-}\to\Omega_{b}^{-}}$  & $-0.079$  & $g_{3,A}^{\Omega_{bb}^{-}\to\Omega_{b}^{-}}$  & $0.250$ \tabularnewline
			\hline
		\end{tabular}
	\end{table}
	
	\begin{table}
		\caption{Values of form factors $f_{i}^{T}$ and $g_{i}^{T}$ at $q^{2}=0$
			for $b\to s$ process in $bb$ sector. The left (right) half of the
			table corresponds to a scalar diquark (an axial-vector diquark) involved
			case. $f_{1}^{T}$ and $g_{1}^{T}$ are obtained by Eqs.~(\ref{eq:f1T}) and (\ref{eq:g1T}) respectively.}
		\label{Tab:fTgT_bb} %
		\begin{tabular}{c|r|c|r||c|r|c|r}
			\hline
			$F$  & $F(0)$  & $F$  & $F(0)$  & $F$  & $F(0)$  & $F$  & $F(0)$ \tabularnewline
			\hline
			$f_{2,S}^{T,\Xi_{bb}\to\Xi_{b}}$  & $0.108$  & $g_{2,S}^{T,\Xi_{bb}\to\Xi_{b}}$  & $0.128$  & $f_{2,A}^{T,\Xi_{bb}\to\Xi_{b}}$  & $-0.066$  & $g_{2,A}^{T,\Xi_{bb}\to\Xi_{b}}$  & $-0.049$ \tabularnewline
			$f_{3,S}^{T,\Xi_{bb}\to\Xi_{b}}$  & $0.091$  & $g_{3,S}^{T,\Xi_{bb}\to\Xi_{b}}$  & $0.156$  & $f_{3,A}^{T,\Xi_{bb}\to\Xi_{b}}$  & $0.134$  & $g_{3,A}^{T,\Xi_{bb}\to\Xi_{b}}$  & $0.032$ \tabularnewline
			\hline
			$f_{2,S}^{T,\Xi_{bb}\to\Xi_{b}^{\prime}}$  & $0.117$  & $g_{2,S}^{T,\Xi_{bb}\to\Xi_{b}^{\prime}}$  & $0.127$  & $f_{2,A}^{T,\Xi_{bb}\to\Xi_{b}^{\prime}}$  & $-0.068$  & $g_{2,A}^{T,\Xi_{bb}\to\Xi_{b}^{\prime}}$  & $-0.049$ \tabularnewline
			$f_{3,S}^{T,\Xi_{bb}\to\Xi_{b}^{\prime}}$  & $0.091$  & $g_{3,S}^{T,\Xi_{bb}\to\Xi_{b}^{\prime}}$  & $0.198$  & $f_{3,A}^{T,\Xi_{bb}\to\Xi_{b}^{\prime}}$  & $0.134$  & $g_{3,A}^{T,\Xi_{bb}\to\Xi_{b}^{\prime}}$  & $0.026$ \tabularnewline
			\hline
			$f_{2,S}^{T,\Omega_{bb}^{-}\to\Omega_{b}^{-}}$  & $0.112$  & $g_{2,S}^{T,\Omega_{bb}^{-}\to\Omega_{b}^{-}}$  & $0.123$  & $f_{2,A}^{T,\Omega_{bb}^{-}\to\Omega_{b}^{-}}$  & $-0.065$  & $g_{2,A}^{T,\Omega_{bb}^{-}\to\Omega_{b}^{-}}$  & $-0.047$ \tabularnewline
			$f_{3,S}^{T,\Omega_{bb}^{-}\to\Omega_{b}^{-}}$  & $0.088$  & $g_{3,S}^{T,\Omega_{bb}^{-}\to\Omega_{b}^{-}}$  & $0.186$  & $f_{3,A}^{T,\Omega_{bb}^{-}\to\Omega_{b}^{-}}$  & $0.130$  & $g_{3,A}^{T,\Omega_{bb}^{-}\to\Omega_{b}^{-}}$  & $0.027$ \tabularnewline
			\hline
		\end{tabular}
	\end{table}
	
	\begin{table}
		\caption{Same as Table \ref{Tab:fg_bb} but for $b\to s$ process in $bc$
			sector. $bc^{\prime}$ sector has the same form factors.}
		\label{Tab:fg_bc} %
		\begin{tabular}{c|r|c|r||c|r|c|r}
			\hline
			$F$  & $F(0)$  & $F$  & $F(0)$  & $F$  & $F(0)$  & $F$  & $F(0)$ \tabularnewline
			\hline
			$f_{1,S}^{\Xi_{bc}\to\Xi_{c}}$  & $0.203$  & $g_{1,S}^{\Xi_{bc}\to\Xi_{c}}$  & $0.167$  & $f_{1,A}^{\Xi_{bc}\to\Xi_{c}}$  & $0.185$  & $g_{1,A}^{\Xi_{bc}\to\Xi_{c}}$  & $-0.033$ \tabularnewline
			$f_{2,S}^{\Xi_{bc}\to\Xi_{c}}$  & $-0.079$  & $g_{2,S}^{\Xi_{bc}\to\Xi_{c}}$  & $0.097$  & $f_{2,A}^{\Xi_{bc}\to\Xi_{c}}$  & $0.203$  & $g_{2,A}^{\Xi_{bc}\to\Xi_{c}}$  & $-0.068$ \tabularnewline
			$f_{3,S}^{\Xi_{bc}\to\Xi_{c}}$  & $0.015$  & $g_{3,S}^{\Xi_{bc}\to\Xi_{c}}$  & $-0.329$  & $f_{3,A}^{\Xi_{bc}\to\Xi_{c}}$  & $-0.109$  & $g_{3,A}^{\Xi_{bc}\to\Xi_{c}}$  & $0.166$ \tabularnewline
			\hline
			$f_{1,S}^{\Xi_{bc}\to\Xi_{c}^{\prime}}$  & $0.204$  & $g_{1,S}^{\Xi_{bc}\to\Xi_{c}^{\prime}}$  & $0.174$  & $f_{1,A}^{\Xi_{bc}\to\Xi_{c}^{\prime}}$  & $0.186$  & $g_{1,A}^{\Xi_{bc}\to\Xi_{c}^{\prime}}$  & $-0.035$ \tabularnewline
			$f_{2,S}^{\Xi_{bc}\to\Xi_{c}^{\prime}}$  & $-0.090$  & $g_{2,S}^{\Xi_{bc}\to\Xi_{c}^{\prime}}$  & $0.074$  & $f_{2,A}^{\Xi_{bc}\to\Xi_{c}^{\prime}}$  & $0.205$  & $g_{2,A}^{\Xi_{bc}\to\Xi_{c}^{\prime}}$  & $-0.063$ \tabularnewline
			$f_{3,S}^{\Xi_{bc}\to\Xi_{c}^{\prime}}$  & $0.007$  & $g_{3,S}^{\Xi_{bc}\to\Xi_{c}^{\prime}}$  & $-0.300$  & $f_{3,A}^{\Xi_{bc}\to\Xi_{c}^{\prime}}$  & $-0.116$  & $g_{3,A}^{\Xi_{bc}\to\Xi_{c}^{\prime}}$  & $0.164$ \tabularnewline
			\hline
			$f_{1,S}^{\Omega_{bc}^{0}\to\Omega_{c}^{0}}$  & $0.192$  & $g_{1,S}^{\Omega_{bc}^{0}\to\Omega_{c}^{0}}$  & $0.165$  & $f_{1,A}^{\Omega_{bc}^{0}\to\Omega_{c}^{0}}$  & $0.177$  & $g_{1,A}^{\Omega_{bc}^{0}\to\Omega_{c}^{0}}$  & $-0.033$ \tabularnewline
			$f_{2,S}^{\Omega_{bc}^{0}\to\Omega_{c}^{0}}$  & $-0.091$  & $g_{2,S}^{\Omega_{bc}^{0}\to\Omega_{c}^{0}}$  & $0.064$  & $f_{2,A}^{\Omega_{bc}^{0}\to\Omega_{c}^{0}}$  & $0.194$  & $g_{2,A}^{\Omega_{bc}^{0}\to\Omega_{c}^{0}}$  & $-0.061$ \tabularnewline
			$f_{3,S}^{\Omega_{bc}^{0}\to\Omega_{c}^{0}}$  & $0.004$  & $g_{3,S}^{\Omega_{bc}^{0}\to\Omega_{c}^{0}}$  & $-0.288$  & $f_{3,A}^{\Omega_{bc}^{0}\to\Omega_{c}^{0}}$  & $-0.112$  & $g_{3,A}^{\Omega_{bc}^{0}\to\Omega_{c}^{0}}$  & $0.163$ \tabularnewline
			\hline
		\end{tabular}
	\end{table}
	
	\begin{table}
		\caption{Same as Table \ref{Tab:fTgT_bb} but for $b\to s$ process in $bc$
			sector. $bc^{\prime}$ sector has the same form factors.}
		\label{Tab:fTgT_bc} %
		\begin{tabular}{c|r|c|r||c|r|c|r}
			\hline
			$F$  & $F(0)$  & $F$  & $F(0)$  & $F$  & $F(0)$  & $F$  & $F(0)$ \tabularnewline
			\hline
			$f_{2,S}^{T,\Xi_{bc}\to\Xi_{c}}$  & $0.160$  & $g_{2,S}^{T,\Xi_{bc}\to\Xi_{c}}$  & $0.202$  & $f_{2,A}^{T,\Xi_{bc}\to\Xi_{c}}$  & $-0.070$  & $g_{2,A}^{T,\Xi_{bc}\to\Xi_{c}}$  & $-0.072$ \tabularnewline
			$f_{3,S}^{T,\Xi_{bc}\to\Xi_{c}}$  & $0.085$  & $g_{3,S}^{T,\Xi_{bc}\to\Xi_{c}}$  & $-0.021$  & $f_{3,A}^{T,\Xi_{bc}\to\Xi_{c}}$  & $0.172$  & $g_{3,A}^{T,\Xi_{bc}\to\Xi_{c}}$  & $0.068$ \tabularnewline
			\hline
			$f_{2,S}^{T,\Xi_{bc}\to\Xi_{c}^{\prime}}$  & $0.169$  & $g_{2,S}^{T,\Xi_{bc}\to\Xi_{c}^{\prime}}$  & $0.200$  & $f_{2,A}^{T,\Xi_{bc}\to\Xi_{c}^{\prime}}$  & $-0.071$  & $g_{2,A}^{T,\Xi_{bc}\to\Xi_{c}^{\prime}}$  & $-0.072$ \tabularnewline
			$f_{3,S}^{T,\Xi_{bc}\to\Xi_{c}^{\prime}}$  & $0.083$  & $g_{3,S}^{T,\Xi_{bc}\to\Xi_{c}^{\prime}}$  & $-0.006$  & $f_{3,A}^{T,\Xi_{bc}\to\Xi_{c}^{\prime}}$  & $0.170$  & $g_{3,A}^{T,\Xi_{bc}\to\Xi_{c}^{\prime}}$  & $0.068$ \tabularnewline
			\hline
			$f_{2,S}^{T,\Omega_{bc}^{0}\to\Omega_{c}^{0}}$  & $0.159$  & $g_{2,S}^{T,\Omega_{bc}^{0}\to\Omega_{c}^{0}}$  & $0.188$  & $f_{2,A}^{T,\Omega_{bc}^{0}\to\Omega_{c}^{0}}$  & $-0.070$  & $g_{2,A}^{T,\Omega_{bc}^{0}\to\Omega_{c}^{0}}$  & $-0.069$ \tabularnewline
			$f_{3,S}^{T,\Omega_{bc}^{0}\to\Omega_{c}^{0}}$  & $0.081$  & $g_{3,S}^{T,\Omega_{bc}^{0}\to\Omega_{c}^{0}}$  & $-0.001$  & $f_{3,A}^{T,\Omega_{bc}^{0}\to\Omega_{c}^{0}}$  & $0.163$  & $g_{3,A}^{T,\Omega_{bc}^{0}\to\Omega_{c}^{0}}$  & $0.067$ \tabularnewline
			\hline
		\end{tabular}
	\end{table}
	
	
	\begin{table}
		\caption{Same as Table \ref{Tab:fg_bb} but for $b\to d$ process.}
		\label{Tab:fg_bb_b2d} %
		\begin{tabular}{c|r|c|r||c|r|c|r}
			\hline
			$F$  & $F(0)$  & $F$  & $F(0)$  & $F$  & $F(0)$  & $F$  & $F(0)$ \tabularnewline
			\hline
			$f_{1,S}^{\Xi_{bb}^{0}\to\Lambda_{b}^{0}}$  & $0.100$  & $g_{1,S}^{\Xi_{bb}^{0}\to\Lambda_{b}^{0}}$  & $0.087$  & $f_{1,A}^{\Xi_{bb}^{0}\to\Lambda_{b}^{0}}$  & $0.098$  & $g_{1,A}^{\Xi_{bb}^{0}\to\Lambda_{b}^{0}}$  & $-0.020$ \tabularnewline
			$f_{2,S}^{\Xi_{bb}^{0}\to\Lambda_{b}^{0}}$  & $-0.136$  & $g_{2,S}^{\Xi_{bb}^{0}\to\Lambda_{b}^{0}}$  & $0.041$  & $f_{2,A}^{\Xi_{bb}^{0}\to\Lambda_{b}^{0}}$  & $0.099$  & $g_{2,A}^{\Xi_{bb}^{0}\to\Lambda_{b}^{0}}$  & $-0.043$ \tabularnewline
			$f_{3,S}^{\Xi_{bb}^{0}\to\Lambda_{b}^{0}}$  & $0.008$  & $g_{3,S}^{\Xi_{bb}^{0}\to\Lambda_{b}^{0}}$  & $-0.298$  & $f_{3,A}^{\Xi_{bb}^{0}\to\Lambda_{b}^{0}}$  & $-0.057$  & $g_{3,A}^{\Xi_{bb}^{0}\to\Lambda_{b}^{0}}$  & $0.191$ \tabularnewline
			\hline
			$f_{1,S}^{\Xi_{bb}^{0,-}\to\Sigma_{b}^{0,-}}$  & $0.102$  & $g_{1,S}^{\Xi_{bb}^{0,-}\to\Sigma_{b}^{0,-}}$  & $0.094$  & $f_{1,A}^{\Xi_{bb}^{0,-}\to\Sigma_{b}^{0,-}}$  & $0.100$  & $g_{1,A}^{\Xi_{bb}^{0,-}\to\Sigma_{b}^{0,-}}$  & $-0.021$ \tabularnewline
			$f_{2,S}^{\Xi_{bb}^{0,-}\to\Sigma_{b}^{0,-}}$  & $-0.150$  & $g_{2,S}^{\Xi_{bb}^{0,-}\to\Sigma_{b}^{0,-}}$  & $0.012$  & $f_{2,A}^{\Xi_{bb}^{0,-}\to\Sigma_{b}^{0,-}}$  & $0.104$  & $g_{2,A}^{\Xi_{bb}^{0,-}\to\Sigma_{b}^{0,-}}$  & $-0.037$ \tabularnewline
			$f_{3,S}^{\Xi_{bb}^{0,-}\to\Sigma_{b}^{0,-}}$  & $-0.004$  & $g_{3,S}^{\Xi_{bb}^{0,-}\to\Sigma_{b}^{0,-}}$  & $-0.222$  & $f_{3,A}^{\Xi_{bb}^{0,-}\to\Sigma_{b}^{0,-}}$  & $-0.070$  & $g_{3,A}^{\Xi_{bb}^{0,-}\to\Sigma_{b}^{0,-}}$  & $0.183$ \tabularnewline
			\hline
			$f_{1,S}^{\Omega_{bb}^{-}\to\Xi_{b}^{-}}$  & $0.098$  & $g_{1,S}^{\Omega_{bb}^{-}\to\Xi_{b}^{-}}$  & $0.086$  & $f_{1,A}^{\Omega_{bb}^{-}\to\Xi_{b}^{-}}$  & $0.095$  & $g_{1,A}^{\Omega_{bb}^{-}\to\Xi_{b}^{-}}$  & $-0.020$ \tabularnewline
			$f_{2,S}^{\Omega_{bb}^{-}\to\Xi_{b}^{-}}$  & $-0.137$  & $g_{2,S}^{\Omega_{bb}^{-}\to\Xi_{b}^{-}}$  & $0.034$  & $f_{2,A}^{\Omega_{bb}^{-}\to\Xi_{b}^{-}}$  & $0.098$  & $g_{2,A}^{\Omega_{bb}^{-}\to\Xi_{b}^{-}}$  & $-0.040$ \tabularnewline
			$f_{3,S}^{\Omega_{bb}^{-}\to\Xi_{b}^{-}}$  & $0.004$  & $g_{3,S}^{\Omega_{bb}^{-}\to\Xi_{b}^{-}}$  & $-0.282$  & $f_{3,A}^{\Omega_{bb}^{-}\to\Xi_{b}^{-}}$  & $-0.059$  & $g_{3,A}^{\Omega_{bb}^{-}\to\Xi_{b}^{-}}$  & $0.187$ \tabularnewline
			\hline
			$f_{1,S}^{\Omega_{bb}^{-}\to\Xi_{b}^{\prime-}}$  & $0.099$  & $g_{1,S}^{\Omega_{bb}^{-}\to\Xi_{b}^{\prime-}}$  & $0.091$  & $f_{1,A}^{\Omega_{bb}^{-}\to\Xi_{b}^{\prime-}}$  & $0.097$  & $g_{1,A}^{\Omega_{bb}^{-}\to\Xi_{b}^{\prime-}}$  & $-0.021$ \tabularnewline
			$f_{2,S}^{\Omega_{bb}^{-}\to\Xi_{b}^{\prime-}}$  & $-0.147$  & $g_{2,S}^{\Omega_{bb}^{-}\to\Xi_{b}^{\prime-}}$  & $0.013$  & $f_{2,A}^{\Omega_{bb}^{-}\to\Xi_{b}^{\prime-}}$  & $0.102$  & $g_{2,A}^{\Omega_{bb}^{-}\to\Xi_{b}^{\prime-}}$  & $-0.036$ \tabularnewline
			$f_{3,S}^{\Omega_{bb}^{-}\to\Xi_{b}^{\prime-}}$  & $-0.005$  & $g_{3,S}^{\Omega_{bb}^{-}\to\Xi_{b}^{\prime-}}$  & $-0.226$  & $f_{3,A}^{\Omega_{bb}^{-}\to\Xi_{b}^{\prime-}}$  & $-0.068$  & $g_{3,A}^{\Omega_{bb}^{-}\to\Xi_{b}^{\prime-}}$  & $0.181$ \tabularnewline
			\hline
		\end{tabular}
	\end{table}
	
	\begin{table}
		\caption{Same as Table \ref{Tab:fTgT_bb} for $b\to d$ process.}
		\label{Tab:fTgT_bb_b2d} %
		\begin{tabular}{c|r|c|r||c|r|c|r}
			\hline
			$F$  & $F(0)$  & $F$  & $F(0)$  & $F$  & $F(0)$  & $F$  & $F(0)$ \tabularnewline
			\hline
			$f_{2,S}^{T,\Xi_{bb}^{0}\to\Lambda_{b}^{0}}$  & $0.075$  & $g_{2,S}^{T,\Xi_{bb}^{0}\to\Lambda_{b}^{0}}$  & $0.091$  & $f_{2,A}^{T,\Xi_{bb}^{0}\to\Lambda_{b}^{0}}$  & $-0.049$  & $g_{2,A}^{T,\Xi_{bb}^{0}\to\Lambda_{b}^{0}}$  & $-0.035$ \tabularnewline
			$f_{3,S}^{T,\Xi_{bb}^{0}\to\Lambda_{b}^{0}}$  & $0.072$  & $g_{3,S}^{T,\Xi_{bb}^{0}\to\Lambda_{b}^{0}}$  & $0.114$  & $f_{3,A}^{T,\Xi_{bb}^{0}\to\Lambda_{b}^{0}}$  & $0.104$  & $g_{3,A}^{T,\Xi_{bb}^{0}\to\Lambda_{b}^{0}}$  & $0.028$ \tabularnewline
			\hline
			$f_{2,S}^{T,\Xi_{bb}^{0,-}\to\Sigma_{b}^{0,-}}$  & $0.083$  & $g_{2,S}^{T,\Xi_{bb}^{0,-}\to\Sigma_{b}^{0,-}}$  & $0.090$  & $f_{2,A}^{T,\Xi_{bb}^{0,-}\to\Sigma_{b}^{0,-}}$  & $-0.051$  & $g_{2,A}^{T,\Xi_{bb}^{0,-}\to\Sigma_{b}^{0,-}}$  & $-0.035$ \tabularnewline
			$f_{3,S}^{T,\Xi_{bb}^{0,-}\to\Sigma_{b}^{0,-}}$  & $0.072$  & $g_{3,S}^{T,\Xi_{bb}^{0,-}\to\Sigma_{b}^{0,-}}$  & $0.154$  & $f_{3,A}^{T,\Xi_{bb}^{0,-}\to\Sigma_{b}^{0,-}}$  & $0.104$  & $g_{3,A}^{T,\Xi_{bb}^{0,-}\to\Sigma_{b}^{0,-}}$  & $0.023$ \tabularnewline
			\hline
			$f_{2,S}^{T,\Omega_{bb}^{-}\to\Xi_{b}^{-}}$  & $0.074$  & $g_{2,S}^{T,\Omega_{bb}^{-}\to\Xi_{b}^{-}}$  & $0.088$  & $f_{2,A}^{T,\Omega_{bb}^{-}\to\Xi_{b}^{-}}$  & $-0.048$  & $g_{2,A}^{T,\Omega_{bb}^{-}\to\Xi_{b}^{-}}$  & $-0.034$ \tabularnewline
			$f_{3,S}^{T,\Omega_{bb}^{-}\to\Xi_{b}^{-}}$  & $0.069$  & $g_{3,S}^{T,\Omega_{bb}^{-}\to\Xi_{b}^{-}}$  & $0.119$  & $f_{3,A}^{T,\Omega_{bb}^{-}\to\Xi_{b}^{-}}$  & $0.100$  & $g_{3,A}^{T,\Omega_{bb}^{-}\to\Xi_{b}^{-}}$  & $0.026$ \tabularnewline
			\hline
			$f_{2,S}^{T,\Omega_{bb}^{-}\to\Xi_{b}^{\prime-}}$  & $0.080$  & $g_{2,S}^{T,\Omega_{bb}^{-}\to\Xi_{b}^{\prime-}}$  & $0.087$  & $f_{2,A}^{T,\Omega_{bb}^{-}\to\Xi_{b}^{\prime-}}$  & $-0.049$  & $g_{2,A}^{T,\Omega_{bb}^{-}\to\Xi_{b}^{\prime-}}$  & $-0.034$ \tabularnewline
			$f_{3,S}^{T,\Omega_{bb}^{-}\to\Xi_{b}^{\prime-}}$  & $0.069$  & $g_{3,S}^{T,\Omega_{bb}^{-}\to\Xi_{b}^{\prime-}}$  & $0.148$  & $f_{3,A}^{T,\Omega_{bb}^{-}\to\Xi_{b}^{\prime-}}$  & $0.101$  & $g_{3,A}^{T,\Omega_{bb}^{-}\to\Xi_{b}^{\prime-}}$  & $0.023$ \tabularnewline
			\hline
		\end{tabular}
	\end{table}
	
	\begin{table}
		\caption{Same as Table \ref{Tab:fg_bc} but for $b\to d$ process.}
		\label{Tab:fg_bc_b2d} %
		\begin{tabular}{c|r|c|r||c|r|c|r}
			\hline
			$F$  & $F(0)$  & $F$  & $F(0)$  & $F$  & $F(0)$  & $F$  & $F(0)$ \tabularnewline
			\hline
			$f_{1,S}^{\Xi_{bc}^{+}\to\Lambda_{c}^{+}}$  & $0.143$  & $g_{1,S}^{\Xi_{bc}^{+}\to\Lambda_{c}^{+}}$  & $0.117$  & $f_{1,A}^{\Xi_{bc}^{+}\to\Lambda_{c}^{+}}$  & $0.130$  & $g_{1,A}^{\Xi_{bc}^{+}\to\Lambda_{c}^{+}}$  & $-0.020$ \tabularnewline
			$f_{2,S}^{\Xi_{bc}^{+}\to\Lambda_{c}^{+}}$  & $-0.055$  & $g_{2,S}^{\Xi_{bc}^{+}\to\Lambda_{c}^{+}}$  & $0.070$  & $f_{2,A}^{\Xi_{bc}^{+}\to\Lambda_{c}^{+}}$  & $0.149$  & $g_{2,A}^{\Xi_{bc}^{+}\to\Lambda_{c}^{+}}$  & $-0.054$ \tabularnewline
			$f_{3,S}^{\Xi_{bc}^{+}\to\Lambda_{c}^{+}}$  & $0.009$  & $g_{3,S}^{\Xi_{bc}^{+}\to\Lambda_{c}^{+}}$  & $-0.224$  & $f_{3,A}^{\Xi_{bc}^{+}\to\Lambda_{c}^{+}}$  & $-0.087$  & $g_{3,A}^{\Xi_{bc}^{+}\to\Lambda_{c}^{+}}$  & $0.121$ \tabularnewline
			\hline
			$f_{1,S}^{\Xi_{bc}^{+,0}\to\Sigma_{c}^{+,0}}$  & $0.143$  & $g_{1,S}^{\Xi_{bc}^{+,0}\to\Sigma_{c}^{+,0}}$  & $0.123$  & $f_{1,A}^{\Xi_{bc}^{+,0}\to\Sigma_{c}^{+,0}}$  & $0.130$  & $g_{1,A}^{\Xi_{bc}^{+,0}\to\Sigma_{c}^{+,0}}$  & $-0.021$ \tabularnewline
			$f_{2,S}^{\Xi_{bc}^{+,0}\to\Sigma_{c}^{+,0}}$  & $-0.067$  & $g_{2,S}^{\Xi_{bc}^{+,0}\to\Sigma_{c}^{+,0}}$  & $0.046$  & $f_{2,A}^{\Xi_{bc}^{+,0}\to\Sigma_{c}^{+,0}}$  & $0.150$  & $g_{2,A}^{\Xi_{bc}^{+,0}\to\Sigma_{c}^{+,0}}$  & $-0.050$ \tabularnewline
			$f_{3,S}^{\Xi_{bc}^{+,0}\to\Sigma_{c}^{+,0}}$  & $0.001$  & $g_{3,S}^{\Xi_{bc}^{+,0}\to\Sigma_{c}^{+,0}}$  & $-0.197$  & $f_{3,A}^{\Xi_{bc}^{+,0}\to\Sigma_{c}^{+,0}}$  & $-0.094$  & $g_{3,A}^{\Xi_{bc}^{+,0}\to\Sigma_{c}^{+,0}}$  & $0.121$ \tabularnewline
			\hline
			$f_{1,S}^{\Omega_{bc}^{0}\to\Xi_{c}^{0}}$  & $0.133$  & $g_{1,S}^{\Omega_{bc}^{0}\to\Xi_{c}^{0}}$  & $0.111$  & $f_{1,A}^{\Omega_{bc}^{0}\to\Xi_{c}^{0}}$  & $0.122$  & $g_{1,A}^{\Omega_{bc}^{0}\to\Xi_{c}^{0}}$  & $-0.019$ \tabularnewline
			$f_{2,S}^{\Omega_{bc}^{0}\to\Xi_{c}^{0}}$  & $-0.060$  & $g_{2,S}^{\Omega_{bc}^{0}\to\Xi_{c}^{0}}$  & $0.053$  & $f_{2,A}^{\Omega_{bc}^{0}\to\Xi_{c}^{0}}$  & $0.139$  & $g_{2,A}^{\Omega_{bc}^{0}\to\Xi_{c}^{0}}$  & $-0.049$ \tabularnewline
			$f_{3,S}^{\Omega_{bc}^{0}\to\Xi_{c}^{0}}$  & $0.003$  & $g_{3,S}^{\Omega_{bc}^{0}\to\Xi_{c}^{0}}$  & $-0.204$  & $f_{3,A}^{\Omega_{bc}^{0}\to\Xi_{c}^{0}}$  & $-0.085$  & $g_{3,A}^{\Omega_{bc}^{0}\to\Xi_{c}^{0}}$  & $0.118$ \tabularnewline
			\hline
			$f_{1,S}^{\Omega_{bc}^{0}\to\Xi_{c}^{\prime0}}$  & $0.133$  & $g_{1,S}^{\Omega_{bc}^{0}\to\Xi_{c}^{\prime0}}$  & $0.116$  & $f_{1,A}^{\Omega_{bc}^{0}\to\Xi_{c}^{\prime0}}$  & $0.122$  & $g_{1,A}^{\Omega_{bc}^{0}\to\Xi_{c}^{\prime0}}$  & $-0.020$ \tabularnewline
			$f_{2,S}^{\Omega_{bc}^{0}\to\Xi_{c}^{\prime0}}$  & $-0.067$  & $g_{2,S}^{\Omega_{bc}^{0}\to\Xi_{c}^{\prime0}}$  & $0.038$  & $f_{2,A}^{\Omega_{bc}^{0}\to\Xi_{c}^{\prime0}}$  & $0.140$  & $g_{2,A}^{\Omega_{bc}^{0}\to\Xi_{c}^{\prime0}}$  & $-0.047$ \tabularnewline
			$f_{3,S}^{\Omega_{bc}^{0}\to\Xi_{c}^{\prime0}}$  & $-0.001$  & $g_{3,S}^{\Omega_{bc}^{0}\to\Xi_{c}^{\prime0}}$  & $-0.185$  & $f_{3,A}^{\Omega_{bc}^{0}\to\Xi_{c}^{\prime0}}$  & $-0.089$  & $g_{3,A}^{\Omega_{bc}^{0}\to\Xi_{c}^{\prime0}}$  & $0.118$ \tabularnewline
			\hline
		\end{tabular}
	\end{table}
	
	\begin{table}
		\caption{Same as Table \ref{Tab:fTgT_bc} but for $b\to d$ process.}
		\label{Tab:fTgT_bc_b2d} %
		\begin{tabular}{c|r|c|r||c|r|c|r}
			\hline
			$F$  & $F(0)$  & $F$  & $F(0)$  & $F$  & $F(0)$  & $F$  & $F(0)$ \tabularnewline
			\hline
			$f_{2,S}^{T,\Xi_{bc}^{+}\to\Lambda_{c}^{+}}$  & $0.110$  & $g_{2,S}^{T,\Xi_{bc}^{+}\to\Lambda_{c}^{+}}$  & $0.142$  & $f_{2,A}^{T,\Xi_{bc}^{+}\to\Lambda_{c}^{+}}$  & $-0.052$  & $g_{2,A}^{T,\Xi_{bc}^{+}\to\Lambda_{c}^{+}}$  & $-0.052$ \tabularnewline
			$f_{3,S}^{T,\Xi_{bc}^{+}\to\Lambda_{c}^{+}}$  & $0.068$  & $g_{3,S}^{T,\Xi_{bc}^{+}\to\Lambda_{c}^{+}}$  & $-0.010$  & $f_{3,A}^{T,\Xi_{bc}^{+}\to\Lambda_{c}^{+}}$  & $0.133$  & $g_{3,A}^{T,\Xi_{bc}^{+}\to\Lambda_{c}^{+}}$  & $0.055$ \tabularnewline
			\hline
			$f_{2,S}^{T,\Xi_{bc}^{+,0}\to\Sigma_{c}^{+,0}}$  & $0.119$  & $g_{2,S}^{T,\Xi_{bc}^{+,0}\to\Sigma_{c}^{+,0}}$  & $0.140$  & $f_{2,A}^{T,\Xi_{bc}^{+,0}\to\Sigma_{c}^{+,0}}$  & $-0.053$  & $g_{2,A}^{T,\Xi_{bc}^{+,0}\to\Sigma_{c}^{+,0}}$  & $-0.052$ \tabularnewline
			$f_{3,S}^{T,\Xi_{bc}^{+,0}\to\Sigma_{c}^{+,0}}$  & $0.064$  & $g_{3,S}^{T,\Xi_{bc}^{+,0}\to\Sigma_{c}^{+,0}}$  & $0.006$  & $f_{3,A}^{T,\Xi_{bc}^{+,0}\to\Sigma_{c}^{+,0}}$  & $0.130$  & $g_{3,A}^{T,\Xi_{bc}^{+,0}\to\Sigma_{c}^{+,0}}$  & $0.055$ \tabularnewline
			\hline
			$f_{2,S}^{T,\Omega_{bc}^{0}\to\Xi_{c}^{0}}$  & $0.105$  & $g_{2,S}^{T,\Omega_{bc}^{0}\to\Xi_{c}^{0}}$  & $0.131$  & $f_{2,A}^{T,\Omega_{bc}^{0}\to\Xi_{c}^{0}}$  & $-0.050$  & $g_{2,A}^{T,\Omega_{bc}^{0}\to\Xi_{c}^{0}}$  & $-0.049$ \tabularnewline
			$f_{3,S}^{T,\Omega_{bc}^{0}\to\Xi_{c}^{0}}$  & $0.064$  & $g_{3,S}^{T,\Omega_{bc}^{0}\to\Xi_{c}^{0}}$  & $-0.001$  & $f_{3,A}^{T,\Omega_{bc}^{0}\to\Xi_{c}^{0}}$  & $0.124$  & $g_{3,A}^{T,\Omega_{bc}^{0}\to\Xi_{c}^{0}}$  & $0.053$ \tabularnewline
			\hline
			$f_{2,S}^{T,\Omega_{bc}^{0}\to\Xi_{c}^{\prime0}}$  & $0.110$  & $g_{2,S}^{T,\Omega_{bc}^{0}\to\Xi_{c}^{\prime0}}$  & $0.129$  & $f_{2,A}^{T,\Omega_{bc}^{0}\to\Xi_{c}^{\prime0}}$  & $-0.051$  & $g_{2,A}^{T,\Omega_{bc}^{0}\to\Xi_{c}^{\prime0}}$  & $-0.049$ \tabularnewline
			$f_{3,S}^{T,\Omega_{bc}^{0}\to\Xi_{c}^{\prime0}}$  & $0.062$  & $g_{3,S}^{T,\Omega_{bc}^{0}\to\Xi_{c}^{\prime0}}$  & $0.010$  & $f_{3,A}^{T,\Omega_{bc}^{0}\to\Xi_{c}^{\prime0}}$  & $0.123$  & $g_{3,A}^{T,\Omega_{bc}^{0}\to\Xi_{c}^{\prime0}}$  & $0.053$ \tabularnewline
			\hline
		\end{tabular}
	\end{table}
	
	\subsection{Results for phenomenological observables}
	
	The decay widths are shown in Tables~\ref{Tab:width_bb} to \ref{Tab:width_bcp} for $b\to s$ process and Tables~\ref{Tab:width_bb_b2d} to \ref{Tab:width_bcp_b2d} for $b\to d$ process. Some comments are given in order.
	\begin{itemize}
		\item Since there exist uncertainties in the lifetimes of the parent baryons,
		there may exist small fluctuations in the results for branching ratios.
		\item It can be seen from these tables that, the decay widths are very close
		to each other for $l=e/\mu$ cases, while it is roughly one order of magnitude
		smaller for $l=\tau$ case. This can be attributed to the much smaller
		phase space for $l=\tau$ case.
		\item Most of the branching ratios are $10^{-8}\sim10^{-7}$ for $b\to s$ process and $10^{-9}\sim10^{-8}$ for $b\to d$ process,
		which are roughly one order of magnitude smaller than the corresponding mesonic cases. This is because we believe that the lifetime of the doubly heavy baryon is roughly one order of magnitude smaller than that of $B$ meson.
	\end{itemize}

\begin{figure}[!]
		\includegraphics[width=0.45\columnwidth]{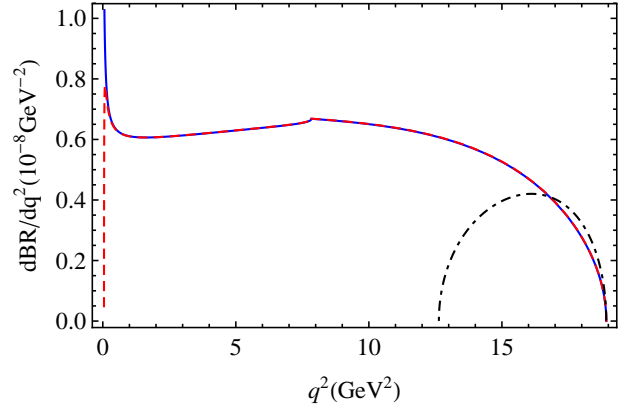} \caption{$d{\cal B}/dq^{2}$ for $\Xi_{bb}^{0}\to \Xi_{b}^{0} l^{+}l^{-}$ with $l=e,\mu,\tau$. The blue solid line, the
			red dashed line and the black dotdashed line correspond to the cases of $l=e,\mu,\tau$, respectively. Here the resonant contributions are not taken into account.}
		\label{fig:dBRdq2_eMuTau}
	\end{figure}

\begin{figure}[!]
		\includegraphics[width=0.47\columnwidth]{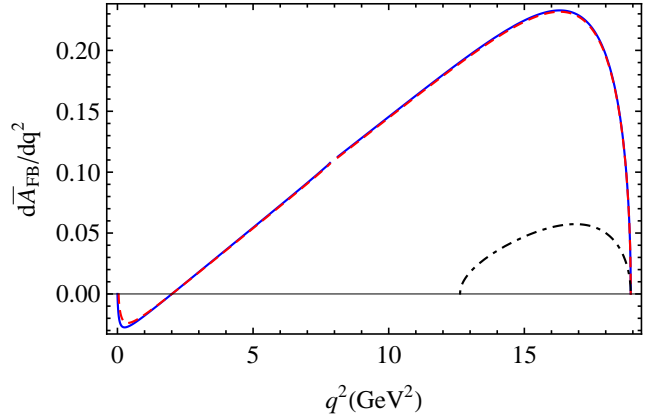} 
		\caption{Same as FIG. \ref{fig:dBRdq2_eMuTau} but for $d\bar{A}_{FB}/dq^{2}$.}
		\label{fig:dAbarFBdq2_eMuTau}
	\end{figure}

	The differential decay widths for $\Xi_{bb}^{0}\to \Xi_{b}^{0} l^{+}l^{-}$ with $l=e,\mu,\tau$ are plotted in Fig.~\ref{fig:dBRdq2_eMuTau}, where the resonant contributions are not taken into account. It can be seen that the curves for $l=e/\mu$ almost coincide with
	each other and the much smaller phase space for $l=\tau$ case can
	be seen clearly. The curves of forward-backward asymmetry (FBA) for $\Xi_{bb}^{0}\to \Xi_{b}^{0} l^{+}l^{-}$ with $l=e,\mu,\tau$ are plotted in Fig.~\ref{fig:dAbarFBdq2_eMuTau}.
	It can be seen from this figure that, the zero-crossing point is around $q^{2}\approx2\,{\rm GeV}^{2}$
	for $l=e/\mu$ cases. The zero-crossing points for other $b\to s$ processes and for $b\to d$ processes can
	be found in Tables~\ref{Tab:AFB_zero_b2s} and \ref{Tab:AFB_zero_b2d} respectively. It can be seen from these tables that these $s_{0}$ roughly range from 2 to 3 GeV$^{2}$.
	
	Following Ref.~\cite{Lu:2011jm}, we now analyse the zero-crossing point $s_{0}$ of FBA which satisfies
	\begin{equation}
	\frac{d\bar{A}_{FB}}{dq^{2}}=\frac{I_{1}}{2(I_{0}-I_{2}/3)}=0
	\end{equation}
	or
	\begin{equation}
	{\rm Re}(C_{9}^{{\rm eff}}(s_{0}))+2\frac{m_{b}M}{s_{0}}C_{7}^{{\rm eff}}{\cal R}(s_{0})=0.
	\end{equation}
	Here ${\cal R}$ is defined by
	\begin{equation}
	{\cal R}\equiv\frac{AD-BC}{2AB}\label{eq:R}
	\end{equation}
	with
	\begin{eqnarray}
	A & = & Mf_{1}-(M+M^{\prime})f_{2},\nonumber \\
	B & = & Mg_{1}+(M-M^{\prime})g_{2},\nonumber \\
	C & = & Mf_{1}^{T}-(M+M^{\prime})f_{2}^{T},\nonumber \\
	D & = & Mg_{1}^{T}+(M-M^{\prime})g_{2}^{T}.\label{eq:ABCD}
	\end{eqnarray}
	
	The meaning of ${\cal R}$ can be seen more clear in $\Lambda_{b}\to\Lambda$
	process with the help of the heavy quark symmetry. In the heavy quark
	symmetry limit, the matrix elements of all the hadronic currents can be parameterized by only two independent form factors~\cite{Manohar:2000dt}
	\begin{equation}
	\langle\Lambda(p_{\Lambda})|\bar{s}\Gamma b|\Lambda_{b}(p_{\Lambda_{b}})\rangle=\bar{u}_{\Lambda}[F_{1}(q^{2})+\slashed vF_{2}(q^{2})]\Gamma u_{\Lambda_{b}},
	\end{equation}
	where $\Gamma$ is the product of Dirac matrices, $v^{\mu}\equiv p_{\Lambda_{b}}^{\mu}/m_{\Lambda_{b}}$
	is the four velocity of $\Lambda_{b}$.
	
	Under the heavy quark symmetry,
	\begin{eqnarray}
	f_{1},g_{1},f_{2}^{T},g_{2}^{T} & \to & F_{1},\nonumber \\
	f_{2},g_{2} & \to & F_{2},\nonumber \\
	f_{1}^{T},g_{1}^{T} & \to & 0,
	\end{eqnarray}
	and ${\cal R}$ is reduced to the following form
	\begin{equation}
	{\cal R}=\frac{F_{1}^{2}}{F_{1}^{2}-F_{2}^{2}},
	\end{equation}
	where we have also neglected the $m_{\Lambda}/m_{\Lambda_{b}}$ term. If we further take into account the fact that $F_{2}\ll F_{1}$ for $\Lambda_{b}\to\Lambda$ process~\cite{Huang:1998ek,Chen:2001sj,Chen:2001ki}, then
	\begin{equation}
	{\cal R}\approx1.
	\end{equation}

	The values of ${\cal R}$ for FCNC processes of doubly heavy baryons can be found in Tables~\ref{Tab:AFB_zero_b2s} and \ref{Tab:AFB_zero_b2d}. It can be seen from these tables that ${\cal R}$ roughly ranges from 0.3 to 0.4 for $bb$ sector, while it lies in the interval of $[0.6,0.7]$ for $bc$ sector.
	
	\begin{table}
		\caption{Decay widths and branching ratios for $b\to s$ process in $bb$ sector.}
		\label{Tab:width_bb} %
		\begin{tabular}{l|c|c|c}
			\hline
			channels  & $\Gamma/\text{~GeV}$  & ${\cal B}$  & $\Gamma_{L}/\Gamma_{T}$ \tabularnewline
			\hline
			$\Xi_{bb}^{0}\to\Xi_{b}^{0}e^{+}e^{-}$  & $1.98\times10^{-19}$  & $1.11\times10^{-7}$  & $3.48$ \tabularnewline
			$\Xi_{bb}^{0}\to\Xi_{b}^{\prime0}e^{+}e^{-}$  & $5.20\times10^{-19}$  & $2.92\times10^{-7}$  & $0.70$ \tabularnewline
			$\Xi_{bb}^{-}\to\Xi_{b}^{-}e^{+}e^{-}$  & $1.97\times10^{-19}$  & $1.11\times10^{-7}$  & $3.49$ \tabularnewline
			$\Xi_{bb}^{-}\to\Xi_{b}^{\prime-}e^{+}e^{-}$  & $5.20\times10^{-19}$  & $2.92\times10^{-7}$  & $0.70$ \tabularnewline
			$\Omega_{bb}^{-}\to\Omega_{b}^{-}e^{+}e^{-}$  & $1.02\times10^{-18}$  & $1.25\times10^{-6}$  & $0.70$ \tabularnewline
			\hline
			$\Xi_{bb}^{0}\to\Xi_{b}^{0}\mu^{+}\mu^{-}$  & $1.92\times10^{-19}$  & $1.08\times10^{-7}$  & $3.95$ \tabularnewline
			$\Xi_{bb}^{0}\to\Xi_{b}^{\prime0}\mu^{+}\mu^{-}$  & $4.47\times10^{-19}$  & $2.52\times10^{-7}$  & $0.91$ \tabularnewline
			$\Xi_{bb}^{-}\to\Xi_{b}^{-}\mu^{+}\mu^{-}$  & $1.91\times10^{-19}$  & $1.08\times10^{-7}$  & $3.96$ \tabularnewline
			$\Xi_{bb}^{-}\to\Xi_{b}^{\prime-}\mu^{+}\mu^{-}$  & $4.47\times10^{-19}$  & $2.52\times10^{-7}$  & $0.91$ \tabularnewline
			$\Omega_{bb}^{-}\to\Omega_{b}^{-}\mu^{+}\mu^{-}$  & $8.85\times10^{-19}$  & $1.08\times10^{-6}$  & $0.90$ \tabularnewline
			\hline
			$\Xi_{bb}^{0}\to\Xi_{b}^{0}\tau^{+}\tau^{-}$  & $3.72\times10^{-20}$  & $2.09\times10^{-8}$  & $6.17$ \tabularnewline
			$\Xi_{bb}^{0}\to\Xi_{b}^{\prime0}\tau^{+}\tau^{-}$  & $4.87\times10^{-20}$  & $2.74\times10^{-8}$  & $1.02$ \tabularnewline
			$\Xi_{bb}^{-}\to\Xi_{b}^{-}\tau^{+}\tau^{-}$  & $3.69\times10^{-20}$  & $2.07\times10^{-8}$  & $6.18$ \tabularnewline
			$\Xi_{bb}^{-}\to\Xi_{b}^{\prime-}\tau^{+}\tau^{-}$  & $4.87\times10^{-20}$  & $2.74\times10^{-8}$  & $1.02$ \tabularnewline
			$\Omega_{bb}^{-}\to\Omega_{b}^{-}\tau^{+}\tau^{-}$  & $1.02\times10^{-19}$  & $1.24\times10^{-7}$  & $1.00$ \tabularnewline
			\hline
		\end{tabular}
	\end{table}
	
	\begin{table}
		\caption{Decay widths and branching ratios for $b\to s$ process in $bc$ sector.}
		\label{Tab:width_bc} %
		\begin{tabular}{l|c|c|c}
			\hline
			channels  & $\Gamma/\text{~GeV}$  & ${\cal B}$  & $\Gamma_{L}/\Gamma_{T}$ \tabularnewline
			\hline
			$\Xi_{bc}^{+}\to\Xi_{c}^{+}e^{+}e^{-}$  & $1.46\times10^{-19}$  & $5.43\times10^{-8}$  & $2.92$ \tabularnewline
			$\Xi_{bc}^{+}\to\Xi_{c}^{\prime+}e^{+}e^{-}$  & $4.54\times10^{-19}$  & $1.69\times10^{-7}$  & $0.68$ \tabularnewline
			$\Xi_{bc}^{0}\to\Xi_{c}^{0}e^{+}e^{-}$  & $1.46\times10^{-19}$  & $2.06\times10^{-8}$  & $2.93$ \tabularnewline
			$\Xi_{bc}^{0}\to\Xi_{c}^{\prime0}e^{+}e^{-}$  & $4.53\times10^{-19}$  & $6.40\times10^{-8}$  & $0.68$ \tabularnewline
			$\Omega_{bc}^{0}\to\Omega_{c}^{0}e^{+}e^{-}$  & $7.42\times10^{-19}$  & $2.48\times10^{-7}$  & $0.68$ \tabularnewline
			\hline
			$\Xi_{bc}^{+}\to\Xi_{c}^{+}\mu^{+}\mu^{-}$  & $1.40\times10^{-19}$  & $5.21\times10^{-8}$  & $3.44$ \tabularnewline
			$\Xi_{bc}^{+}\to\Xi_{c}^{\prime+}\mu^{+}\mu^{-}$  & $3.97\times10^{-19}$  & $1.47\times10^{-7}$  & $0.86$ \tabularnewline
			$\Xi_{bc}^{0}\to\Xi_{c}^{0}\mu^{+}\mu^{-}$  & $1.40\times10^{-19}$  & $1.98\times10^{-8}$  & $3.45$ \tabularnewline
			$\Xi_{bc}^{0}\to\Xi_{c}^{\prime0}\mu^{+}\mu^{-}$  & $3.95\times10^{-19}$  & $5.59\times10^{-8}$  & $0.86$ \tabularnewline
			$\Omega_{bc}^{0}\to\Omega_{c}^{0}\mu^{+}\mu^{-}$  & $6.41\times10^{-19}$  & $2.14\times10^{-7}$  & $0.88$ \tabularnewline
			\hline
			$\Xi_{bc}^{+}\to\Xi_{c}^{+}\tau^{+}\tau^{-}$  & $3.02\times10^{-20}$  & $1.12\times10^{-8}$  & $4.19$ \tabularnewline
			$\Xi_{bc}^{+}\to\Xi_{c}^{\prime+}\tau^{+}\tau^{-}$  & $6.50\times10^{-20}$  & $2.41\times10^{-8}$  & $0.99$ \tabularnewline
			$\Xi_{bc}^{0}\to\Xi_{c}^{0}\tau^{+}\tau^{-}$  & $2.98\times10^{-20}$  & $4.22\times10^{-9}$  & $4.20$ \tabularnewline
			$\Xi_{bc}^{0}\to\Xi_{c}^{\prime0}\tau^{+}\tau^{-}$  & $6.45\times10^{-20}$  & $9.12\times10^{-9}$  & $0.99$ \tabularnewline
			$\Omega_{bc}^{0}\to\Omega_{c}^{0}\tau^{+}\tau^{-}$  & $9.12\times10^{-20}$  & $3.05\times10^{-8}$  & $0.99$ \tabularnewline
			\hline
		\end{tabular}
	\end{table}
	
	\begin{table}
		\caption{Decay widths and branching ratios for $b\to s$ process in $bc^{\prime}$ sector.}
		\label{Tab:width_bcp} %
		\begin{tabular}{l|c|c|c}
			\hline
			channels  & $\Gamma/\text{~GeV}$  & ${\cal B}$  & $\Gamma_{L}/\Gamma_{T}$ \tabularnewline
			\hline
			$\Xi_{bc}^{\prime+}\to\Xi_{c}^{+}e^{+}e^{-}$  & $1.93\times10^{-19}$  & $7.16\times10^{-8}$  & $0.58$ \tabularnewline
			$\Xi_{bc}^{\prime+}\to\Xi_{c}^{\prime+}e^{+}e^{-}$  & $1.27\times10^{-19}$  & $4.70\times10^{-8}$  & $3.16$ \tabularnewline
			$\Xi_{bc}^{\prime0}\to\Xi_{c}^{0}e^{+}e^{-}$  & $1.92\times10^{-19}$  & $2.72\times10^{-8}$  & $0.58$ \tabularnewline
			$\Xi_{bc}^{\prime0}\to\Xi_{c}^{\prime0}e^{+}e^{-}$  & $1.26\times10^{-19}$  & $1.79\times10^{-8}$  & $3.16$ \tabularnewline
			$\Omega_{bc}^{\prime0}\to\Omega_{c}^{0}e^{+}e^{-}$  & $2.11\times10^{-19}$  & $7.05\times10^{-8}$  & $3.34$ \tabularnewline
			\hline
			$\Xi_{bc}^{\prime+}\to\Xi_{c}^{+}\mu^{+}\mu^{-}$  & $1.69\times10^{-19}$  & $6.27\times10^{-8}$  & $0.71$ \tabularnewline
			$\Xi_{bc}^{\prime+}\to\Xi_{c}^{\prime+}\mu^{+}\mu^{-}$  & $1.21\times10^{-19}$  & $4.48\times10^{-8}$  & $3.87$ \tabularnewline
			$\Xi_{bc}^{\prime0}\to\Xi_{c}^{0}\mu^{+}\mu^{-}$  & $1.68\times10^{-19}$  & $2.38\times10^{-8}$  & $0.71$ \tabularnewline
			$\Xi_{bc}^{\prime0}\to\Xi_{c}^{\prime0}\mu^{+}\mu^{-}$  & $1.20\times10^{-19}$  & $1.70\times10^{-8}$  & $3.88$ \tabularnewline
			$\Omega_{bc}^{\prime0}\to\Omega_{c}^{0}\mu^{+}\mu^{-}$  & $2.01\times10^{-19}$  & $6.71\times10^{-8}$  & $4.15$ \tabularnewline
			\hline
			$\Xi_{bc}^{\prime+}\to\Xi_{c}^{+}\tau^{+}\tau^{-}$  & $3.27\times10^{-20}$  & $1.21\times10^{-8}$  & $0.71$ \tabularnewline
			$\Xi_{bc}^{\prime+}\to\Xi_{c}^{\prime+}\tau^{+}\tau^{-}$  & $2.03\times10^{-20}$  & $7.53\times10^{-9}$  & $4.56$ \tabularnewline
			$\Xi_{bc}^{\prime0}\to\Xi_{c}^{0}\tau^{+}\tau^{-}$  & $3.23\times10^{-20}$  & $4.57\times10^{-9}$  & $0.71$ \tabularnewline
			$\Xi_{bc}^{\prime0}\to\Xi_{c}^{\prime0}\tau^{+}\tau^{-}$  & $2.01\times10^{-20}$  & $2.85\times10^{-9}$  & $4.56$ \tabularnewline
			$\Omega_{bc}^{\prime0}\to\Omega_{c}^{0}\tau^{+}\tau^{-}$  & $2.91\times10^{-20}$  & $9.74\times10^{-9}$  & $4.85$ \tabularnewline
			\hline
		\end{tabular}
	\end{table}
	
	
	\begin{table}
		\caption{Decay widths and branching ratios for $b\to d$ process in $bb$ sector.}
		\label{Tab:width_bb_b2d} %
		\begin{tabular}{l|c|c|c}
			\hline
			channels  & $\Gamma/\text{~GeV}$  & ${\cal B}$  & $\Gamma_{L}/\Gamma_{T}$ \tabularnewline
			\hline
			$\Xi_{bb}^{0}\to\Lambda_{b}^{0}e^{+}e^{-}$  & $6.46\times10^{-21}$  & $3.63\times10^{-9}$  & $3.22$ \tabularnewline
			$\Xi_{bb}^{0}\to\Sigma_{b}^{0}e^{+}e^{-}$  & $1.60\times10^{-20}$  & $9.00\times10^{-9}$  & $0.70$ \tabularnewline
			$\Xi_{bb}^{-}\to\Sigma_{b}^{-}e^{+}e^{-}$  & $3.19\times10^{-20}$  & $1.79\times10^{-8}$  & $0.70$ \tabularnewline
			$\Omega_{bb}^{-}\to\Xi_{b}^{-}e^{+}e^{-}$  & $5.71\times10^{-21}$  & $6.94\times10^{-9}$  & $3.36$ \tabularnewline
			$\Omega_{bb}^{-}\to\Xi_{b}^{\prime-}e^{+}e^{-}$  & $1.54\times10^{-20}$  & $1.88\times10^{-8}$  & $0.70$ \tabularnewline
			\hline
			$\Xi_{bb}^{0}\to\Lambda_{b}^{0}\mu^{+}\mu^{-}$  & $6.32\times10^{-21}$  & $3.55\times10^{-9}$  & $3.51$ \tabularnewline
			$\Xi_{bb}^{0}\to\Sigma_{b}^{0}\mu^{+}\mu^{-}$  & $1.41\times10^{-20}$  & $7.94\times10^{-9}$  & $0.88$ \tabularnewline
			$\Xi_{bb}^{-}\to\Sigma_{b}^{-}\mu^{+}\mu^{-}$  & $2.81\times10^{-20}$  & $1.58\times10^{-8}$  & $0.88$ \tabularnewline
			$\Omega_{bb}^{-}\to\Xi_{b}^{-}\mu^{+}\mu^{-}$  & $5.58\times10^{-21}$  & $6.78\times10^{-9}$  & $3.70$ \tabularnewline
			$\Omega_{bb}^{-}\to\Xi_{b}^{\prime-}\mu^{+}\mu^{-}$  & $1.36\times10^{-20}$  & $1.66\times10^{-8}$  & $0.87$ \tabularnewline
			\hline
			$\Xi_{bb}^{0}\to\Lambda_{b}^{0}\tau^{+}\tau^{-}$  & $1.75\times10^{-21}$  & $9.86\times10^{-10}$  & $5.59$ \tabularnewline
			$\Xi_{bb}^{0}\to\Sigma_{b}^{0}\tau^{+}\tau^{-}$  & $2.10\times10^{-21}$  & $1.18\times10^{-9}$  & $1.01$ \tabularnewline
			$\Xi_{bb}^{-}\to\Sigma_{b}^{-}\tau^{+}\tau^{-}$  & $4.17\times10^{-21}$  & $2.35\times10^{-9}$  & $1.01$ \tabularnewline
			$\Omega_{bb}^{-}\to\Xi_{b}^{-}\tau^{+}\tau^{-}$  & $1.40\times10^{-21}$  & $1.71\times10^{-9}$  & $5.80$ \tabularnewline
			$\Omega_{bb}^{-}\to\Xi_{b}^{\prime-}\tau^{+}\tau^{-}$  & $2.08\times10^{-21}$  & $2.53\times10^{-9}$  & $1.01$ \tabularnewline
			\hline
		\end{tabular}
	\end{table}
	
	\begin{table}
		\caption{Decay widths and branching ratios for $b\to d$ process in $bc$ sector.}
		\label{Tab:width_bc_b2d} %
		\begin{tabular}{l|c|c|c}
			\hline
			channels  & $\Gamma/\text{~GeV}$  & ${\cal B}$  & $\Gamma_{L}/\Gamma_{T}$ \tabularnewline
			\hline
			$\Xi_{bc}^{+}\to\Lambda_{c}^{+}e^{+}e^{-}$  & $4.54\times10^{-21}$  & $1.68\times10^{-9}$  & $2.72$ \tabularnewline
			$\Xi_{bc}^{+}\to\Sigma_{c}^{+}e^{+}e^{-}$  & $1.34\times10^{-20}$  & $4.97\times10^{-9}$  & $0.68$ \tabularnewline
			$\Xi_{bc}^{0}\to\Sigma_{c}^{0}e^{+}e^{-}$  & $2.67\times10^{-20}$  & $3.78\times10^{-9}$  & $0.68$ \tabularnewline
			$\Omega_{bc}^{0}\to\Xi_{c}^{0}e^{+}e^{-}$  & $3.28\times10^{-21}$  & $1.10\times10^{-9}$  & $3.10$ \tabularnewline
			$\Omega_{bc}^{0}\to\Xi_{c}^{\prime0}e^{+}e^{-}$  & $1.04\times10^{-20}$  & $3.47\times10^{-9}$  & $0.68$ \tabularnewline
			\hline
			$\Xi_{bc}^{+}\to\Lambda_{c}^{+}\mu^{+}\mu^{-}$  & $4.40\times10^{-21}$  & $1.63\times10^{-9}$  & $3.05$ \tabularnewline
			$\Xi_{bc}^{+}\to\Sigma_{c}^{+}\mu^{+}\mu^{-}$  & $1.20\times10^{-20}$  & $4.44\times10^{-9}$  & $0.83$ \tabularnewline
			$\Xi_{bc}^{0}\to\Sigma_{c}^{0}\mu^{+}\mu^{-}$  & $2.39\times10^{-20}$  & $3.38\times10^{-9}$  & $0.83$ \tabularnewline
			$\Omega_{bc}^{0}\to\Xi_{c}^{0}\mu^{+}\mu^{-}$  & $3.16\times10^{-21}$  & $1.06\times10^{-9}$  & $3.58$ \tabularnewline
			$\Omega_{bc}^{0}\to\Xi_{c}^{\prime0}\mu^{+}\mu^{-}$  & $9.16\times10^{-21}$  & $3.06\times10^{-9}$  & $0.85$ \tabularnewline
			\hline
			$\Xi_{bc}^{+}\to\Lambda_{c}^{+}\tau^{+}\tau^{-}$  & $1.31\times10^{-21}$  & $4.87\times10^{-10}$  & $3.86$ \tabularnewline
			$\Xi_{bc}^{+}\to\Sigma_{c}^{+}\tau^{+}\tau^{-}$  & $2.54\times10^{-21}$  & $9.43\times10^{-10}$  & $0.98$ \tabularnewline
			$\Xi_{bc}^{0}\to\Sigma_{c}^{0}\tau^{+}\tau^{-}$  & $5.06\times10^{-21}$  & $7.16\times10^{-10}$  & $0.99$ \tabularnewline
			$\Omega_{bc}^{0}\to\Xi_{c}^{0}\tau^{+}\tau^{-}$  & $7.46\times10^{-22}$  & $2.49\times10^{-10}$  & $4.38$ \tabularnewline
			$\Omega_{bc}^{0}\to\Xi_{c}^{\prime0}\tau^{+}\tau^{-}$  & $1.70\times10^{-21}$  & $5.70\times10^{-10}$  & $1.00$ \tabularnewline
			\hline
		\end{tabular}
	\end{table}
	
	\begin{table}
		\caption{Decay widths and branching ratios for $b\to d$ process in $bc^{\prime}$ sector.}
		\label{Tab:width_bcp_b2d} %
		\begin{tabular}{l|c|c|c}
			\hline
			channels  & $\Gamma/\text{~GeV}$  & ${\cal B}$  & $\Gamma_{L}/\Gamma_{T}$ \tabularnewline
			\hline
			$\Xi_{bc}^{\prime+}\to\Lambda_{c}^{+}e^{+}e^{-}$  & $6.61\times10^{-21}$  & $2.45\times10^{-9}$  & $0.54$ \tabularnewline
			$\Xi_{bc}^{\prime+}\to\Sigma_{c}^{+}e^{+}e^{-}$  & $3.55\times10^{-21}$  & $1.32\times10^{-9}$  & $3.17$ \tabularnewline
			$\Xi_{bc}^{\prime0}\to\Sigma_{c}^{0}e^{+}e^{-}$  & $7.09\times10^{-21}$  & $1.00\times10^{-9}$  & $3.17$ \tabularnewline
			$\Omega_{bc}^{\prime0}\to\Xi_{c}^{0}e^{+}e^{-}$  & $4.59\times10^{-21}$  & $1.54\times10^{-9}$  & $0.55$ \tabularnewline
			$\Omega_{bc}^{\prime0}\to\Xi_{c}^{\prime0}e^{+}e^{-}$  & $2.82\times10^{-21}$  & $9.43\times10^{-10}$  & $3.39$ \tabularnewline
			\hline
			$\Xi_{bc}^{\prime+}\to\Lambda_{c}^{+}\mu^{+}\mu^{-}$  & $5.98\times10^{-21}$  & $2.22\times10^{-9}$  & $0.63$ \tabularnewline
			$\Xi_{bc}^{\prime+}\to\Sigma_{c}^{+}\mu^{+}\mu^{-}$  & $3.41\times10^{-21}$  & $1.26\times10^{-9}$  & $3.74$ \tabularnewline
			$\Xi_{bc}^{\prime0}\to\Sigma_{c}^{0}\mu^{+}\mu^{-}$  & $6.81\times10^{-21}$  & $9.62\times10^{-10}$  & $3.75$ \tabularnewline
			$\Omega_{bc}^{\prime0}\to\Xi_{c}^{0}\mu^{+}\mu^{-}$  & $4.06\times10^{-21}$  & $1.36\times10^{-9}$  & $0.67$ \tabularnewline
			$\Omega_{bc}^{\prime0}\to\Xi_{c}^{\prime0}\mu^{+}\mu^{-}$  & $2.71\times10^{-21}$  & $9.05\times10^{-10}$  & $4.06$ \tabularnewline
			\hline
			$\Xi_{bc}^{\prime+}\to\Lambda_{c}^{+}\tau^{+}\tau^{-}$  & $1.60\times10^{-21}$  & $5.95\times10^{-10}$  & $0.65$ \tabularnewline
			$\Xi_{bc}^{\prime+}\to\Sigma_{c}^{+}\tau^{+}\tau^{-}$  & $7.32\times10^{-22}$  & $2.71\times10^{-10}$  & $4.48$ \tabularnewline
			$\Xi_{bc}^{\prime0}\to\Sigma_{c}^{0}\tau^{+}\tau^{-}$  & $1.46\times10^{-21}$  & $2.06\times10^{-10}$  & $4.48$ \tabularnewline
			$\Omega_{bc}^{\prime0}\to\Xi_{c}^{0}\tau^{+}\tau^{-}$  & $8.80\times10^{-22}$  & $2.94\times10^{-10}$  & $0.68$ \tabularnewline
			$\Omega_{bc}^{\prime0}\to\Xi_{c}^{\prime0}\tau^{+}\tau^{-}$  & $5.04\times10^{-22}$  & $1.69\times10^{-10}$  & $4.81$ \tabularnewline
			\hline
		\end{tabular}
	\end{table}
	
	
	\begin{table}
		\caption{Zero-crossing points of $d\bar{A}_{FB}/dq^{2}$ and ${\cal R}$ defined in Eqs.~(\ref{eq:R}) and (\ref{eq:ABCD}) for $b\to s$ process with $l=e/\mu$.}
		\label{Tab:AFB_zero_b2s}%
		\centering
		\begin{tabular}{l|c|c|c|c|c|c|c|c}
			\hline
			channels  & $s_{0}/\text{~GeV}^{2}$ & ${\cal R}(s_{0})$ & channels  & $s_{0}/\text{~GeV}^{2}$ & ${\cal R}(s_{0})$ & channels  & $s_{0}/\text{~GeV}^{2}$  & ${\cal R}(s_{0})$\tabularnewline
			\hline
			$\Xi_{bb}^{0}\to\Xi_{b}^{0}l^{+}l^{-}$  &\multirow{2}{*}{ $2.01$ }& \multirow{2}{*}{$0.30$} & $\Xi_{bc}^{+}\to\Xi_{c}^{+}l^{+}l^{-}$  & \multirow{2}{*}{ $2.80$ } & \multirow{2}{*}{ $0.61$ }& $\Xi_{bc}^{\prime+}\to\Xi_{c}^{+}l^{+}l^{-}$  &\multirow{2}{*}{ $3.12$ }  &\multirow{2}{*}{ $0.68$ }\tabularnewline
			$\Xi_{bb}^{-}\to\Xi_{b}^{-}l^{+}l^{-}$  &  &  & $\Xi_{bc}^{0}\to\Xi_{c}^{0}l^{+}l^{-}$  &  &  & $\Xi_{bc}^{\prime0}\to\Xi_{c}^{0}l^{+}l^{-}$  &  & \tabularnewline
			\hline
			$\Xi_{bb}^{0}\to\Xi_{b}^{\prime0}l^{+}l^{-}$  &\multirow{2}{*}{ $2.88$ } &\multirow{2}{*}{ $0.43$} & $\Xi_{bc}^{+}\to\Xi_{c}^{\prime+}l^{+}l^{-}$  &\multirow{2}{*}{ $3.02$ } &\multirow{2}{*}{ $0.66$ } & $\Xi_{bc}^{\prime+}\to\Xi_{c}^{\prime+}l^{+}l^{-}$  &\multirow{2}{*}{ $2.87$ }  &\multirow{2}{*}{ $0.62$ }\tabularnewline
			$\Xi_{bb}^{-}\to\Xi_{b}^{\prime-}l^{+}l^{-}$  &  &  & $\Xi_{bc}^{0}\to\Xi_{c}^{\prime0}l^{+}l^{-}$  &  &  & $\Xi_{bc}^{\prime0}\to\Xi_{c}^{\prime0}l^{+}l^{-}$  &  & \tabularnewline
			\hline
			$\Omega_{bb}^{-}\to\Omega_{b}^{-}l^{+}l^{-}$  & $2.88$  & $0.42$ & $\Omega_{bc}^{0}\to\Omega_{c}^{0}l^{+}l^{-}$  & $3.00$  & $0.65$ & $\Omega_{bc}^{\prime0}\to\Omega_{c}^{0}l^{+}l^{-}$  & $2.80$  & $0.60$\tabularnewline
			\hline
		\end{tabular}
	\end{table}
	
	\begin{table}
		\caption{Same as \ref{Tab:AFB_zero_b2s} but for $b\to d$ process.}
		\label{Tab:AFB_zero_b2d}%
		\begin{tabular}{l|c|c|c|c|c|c|c|c}
			\hline
			channels  & $s_{0}/\text{~GeV}^{2}$ & ${\cal R}(s_{0})$ & channels  & $s_{0}/\text{~GeV}^{2}$ & ${\cal R}(s_{0})$ & channels  & $s_{0}/\text{~GeV}^{2}$  & ${\cal R}(s_{0})$\tabularnewline
			\hline
			$\Xi_{bb}^{0}\to\Lambda_{b}^{0}l^{+}l^{-}$  & $1.96$ & \multirow{2}{*}{$0.29$} & $\Xi_{bc}^{+}\to\Lambda_{c}^{+}l^{+}l^{-}$  & $2.81$ & $0.61$ & $\Xi_{bc}^{\prime+}\to\Lambda_{c}^{+}l^{+}l^{-}$  & $3.09$  & \multirow{2}{*}{$0.67$}\tabularnewline
			$\Omega_{bb}^{-}\to\Xi_{b}^{-}l^{+}l^{-}$  & $2.00$ &  & $\Omega_{bc}^{0}\to\Xi_{c}^{0}l^{+}l^{-}$  & $2.77$ & $0.60$ & $\Omega_{bc}^{\prime0}\to\Xi_{c}^{0}l^{+}l^{-}$  & $3.11$ &\tabularnewline
			\hline
			$\Xi_{bb}^{0}\to\Sigma_{b}^{0}l^{+}l^{-}$  & \multirow{2}{*}{$2.88$}  & \multirow{2}{*}{$0.43$} & $\Xi_{bc}^{+}\to\Sigma_{c}^{+}l^{+}l^{-}$  & \multirow{2}{*}{$3.02$} & \multirow{2}{*}{$0.66$} & $\Xi_{bc}^{\prime+}\to\Sigma_{c}^{+}l^{+}l^{-}$  & \multirow{2}{*}{$2.91$}  & \multirow{2}{*}{$0.63$}\tabularnewline
			$\Xi_{bb}^{-}\to\Sigma_{b}^{-}l^{+}l^{-}$  & & & $\Xi_{bc}^{0}\to\Sigma_{c}^{0}l^{+}l^{-}$  &  &  & $\Xi_{bc}^{\prime0}\to\Sigma_{c}^{0}l^{+}l^{-}$  &  & \tabularnewline
			\hline
			$\Omega_{bb}^{-}\to\Xi_{b}^{\prime-}l^{+}l^{-}$  & $2.88$  & $0.42$ & $\Omega_{bc}^{0}\to\Xi_{c}^{\prime0}l^{+}l^{-}$  & $3.01$ & $0.65$ & $\Omega_{bc}^{\prime0}\to\Xi_{c}^{\prime0}l^{+}l^{-}$  & $2.84$  & $0.61$\tabularnewline
			\hline
		\end{tabular}
	\end{table}
	
	\subsection{SU(3) analyses}
	
	According to the flavor SU(3) symmetry, there exist the following
	relations among these FCNC processes. These relations can be readily
	derived using the overlapping factors given in Table \ref{Tab:overlapping_factors}.
	For $b\to s$ process, we have
	\begin{eqnarray}
	&  & \Gamma(\Xi_{bb}^{0}\to\Xi_{b}^{0}l^{+}l^{-})=\Gamma(\Xi_{bb}^{-}\to\Xi_{b}^{-}l^{+}l^{-}),\nonumber \\
	&  & \Gamma(\Xi_{bb}^{0}\to\Xi_{b}^{\prime0}l^{+}l^{-})=\Gamma(\Xi_{bb}^{-}\to\Xi_{b}^{\prime-}l^{+}l^{-})=\frac{1}{2}\Gamma(\Omega_{bb}^{-}\to\Omega_{b}^{-}l^{+}l^{-})
	\end{eqnarray}
	for $bb$ sector,
	\begin{eqnarray}
	&  & \Gamma(\Xi_{bc}^{+}\to\Xi_{c}^{+}l^{+}l^{-})=\Gamma(\Xi_{bc}^{0}\to\Xi_{c}^{0}l^{+}l^{-}),\nonumber \\
	&  & \Gamma(\Xi_{bc}^{+}\to\Xi_{c}^{\prime+}l^{+}l^{-})=\Gamma(\Xi_{bc}^{0}\to\Xi_{c}^{\prime0}l^{+}l^{-})=\frac{1}{2}\Gamma(\Omega_{bc}^{0}\to\Omega_{c}^{0}l^{+}l^{-})
	\end{eqnarray}
	for $bc$ sector and
	\begin{eqnarray}
	&  & \Gamma(\Xi_{bc}^{\prime+}\to\Xi_{c}^{+}l^{+}l^{-})=\Gamma(\Xi_{bc}^{\prime0}\to\Xi_{c}^{0}l^{+}l^{-}),\nonumber \\
	&  & \Gamma(\Xi_{bc}^{\prime+}\to\Xi_{c}^{\prime+}l^{+}l^{-})=\Gamma(\Xi_{bc}^{\prime0}\to\Xi_{c}^{\prime0}l^{+}l^{-})=\frac{1}{2}\Gamma(\Omega_{bc}^{\prime0}\to\Omega_{c}^{0}l^{+}l^{-})
	\end{eqnarray}
	for $bc^{\prime}$ sector.
	
	For $b\to d$ process, we have
	\begin{eqnarray}
	&  & \Gamma(\Xi_{bb}^{0}\to\Lambda_{b}^{0}l^{+}l^{-})=\Gamma(\Omega_{bb}^{-}\to\Xi_{b}^{-}l^{+}l^{-}),\nonumber \\
	&  & \Gamma(\Xi_{bb}^{0}\to\Sigma_{b}^{0}l^{+}l^{-})=\frac{1}{2}\Gamma(\Xi_{bb}^{-}\to\Sigma_{b}^{-}l^{+}l^{-})=\Gamma(\Omega_{bb}^{-}\to\Xi_{b}^{\prime0}l^{+}l^{-})
	\end{eqnarray}
	for $bb$ sector,
	\begin{eqnarray}
	&  & \Gamma(\Xi_{bc}^{+}\to\Lambda_{c}^{+}l^{+}l^{-})=\Gamma(\Omega_{bc}^{0}\to\Xi_{c}^{0}l^{+}l^{-}),\nonumber \\
	&  & \Gamma(\Xi_{bc}^{+}\to\Sigma_{c}^{+}l^{+}l^{-})=\frac{1}{2}\Gamma(\Xi_{bc}^{0}\to\Sigma_{c}^{0}l^{+}l^{-})=\Gamma(\Omega_{bc}^{0}\to\Xi_{c}^{\prime0}l^{+}l^{-})
	\end{eqnarray}
	for $bc$ sector and
	\begin{eqnarray}
	&  & \Gamma(\Xi_{bc}^{\prime+}\to\Lambda_{c}^{+}l^{+}l^{-})=\Gamma(\Omega_{bc}^{\prime0}\to\Xi_{c}^{0}l^{+}l^{-}),\nonumber \\
	&  & \Gamma(\Xi_{bc}^{\prime+}\to\Sigma_{c}^{+}l^{+}l^{-})=\frac{1}{2}\Gamma(\Xi_{bc}^{\prime0}\to\Sigma_{c}^{0}l^{+}l^{-})=\Gamma(\Omega_{bc}^{\prime0}\to\Xi_{c}^{\prime0}l^{+}l^{-})
	\end{eqnarray}
	for $bc^{\prime}$ sector.
	
	Quantitative analysis for SU(3) symmetry breaking is given in Tables~\ref{Tab:SU3_breaking_bb}
	to \ref{Tab:SU3_breaking_bcp} for $b\to s$ process and some comments
	on SU(3) symmetry breaking are given as follows.
	\begin{itemize}
		\item SU(3) symmetry breaking is larger for the $Qs$ diquark involved case than that for the $Qu/Qd$ diquark involved case.
		Here $Q=b,c$.
		\item SU(3) symmetry breaking is larger for the $cq$ diquark involved case than that for the $bq$ diquark involved case.
		Here $q=u,d,s$.
		\item SU(3) symmetry breaking is smaller for $l=e/\mu$ cases than that for $l=\tau$ case. This can be attributed to the much smaller phase space for $l=\tau$ case. Smaller phase space is more sensitive to the variation of the masses of baryons in the initial and final
		states.
	\end{itemize}
	\begin{table}
		\caption{Quantitative predictions of SU(3) symmetry breaking for $b\to s$ process in $bb$ sector.}
		\label{Tab:SU3_breaking_bb}%
		\begin{tabular}{l|c|c|c}
			\hline
			channels  & $\Gamma/\text{~GeV}$ (LFQM)  & $\Gamma/\text{~GeV}$ (SU(3))  & $|{\rm LFQM}-{\rm SU(3)}|/{\rm SU}(3)$ \tabularnewline
			\hline
			$\Xi_{bb}^{0}\to\Xi_{b}^{\prime0}e^{+}e^{-}$  & $5.20\times10^{-19}$  & $5.20\times10^{-19}$  & - -\tabularnewline
			$\Xi_{bb}^{-}\to\Xi_{b}^{\prime-}e^{+}e^{-}$  & $5.20\times10^{-19}$  & $5.20\times10^{-19}$  & $0\%$\tabularnewline
			$\Omega_{bb}^{-}\to\Omega_{b}^{-}e^{+}e^{-}$  & $1.02\times10^{-18}$  & $1.04\times10^{-18}$  & $2\%$ \tabularnewline
			\hline
			$\Xi_{bb}^{0}\to\Xi_{b}^{\prime0}\mu^{+}\mu^{-}$  & $4.47\times10^{-19}$  & $4.47\times10^{-19}$  & - - \tabularnewline
			$\Xi_{bb}^{-}\to\Xi_{b}^{\prime-}\mu^{+}\mu^{-}$  & $4.47\times10^{-19}$  & $4.47\times10^{-19}$  & $0\%$ \tabularnewline
			$\Omega_{bb}^{-}\to\Omega_{b}^{-}\mu^{+}\mu^{-}$  & $8.85\times10^{-19}$  & $8.94\times10^{-19}$  & $1\%$ \tabularnewline
			\hline
			$\Xi_{bb}^{0}\to\Xi_{b}^{\prime0}\tau^{+}\tau^{-}$  & $4.87\times10^{-20}$  & $4.87\times10^{-20}$  & - -\tabularnewline
			$\Xi_{bb}^{-}\to\Xi_{b}^{\prime-}\tau^{+}\tau^{-}$  & $4.87\times10^{-20}$  & $4.87\times10^{-20}$  & $0\%$ \tabularnewline
			$\Omega_{bb}^{-}\to\Omega_{b}^{-}\tau^{+}\tau^{-}$  & $1.02\times10^{-19}$  & $9.74\times10^{-20}$  & $5\%$ \tabularnewline
			\hline
		\end{tabular}
	\end{table}
	
	\begin{table}
		\caption{Quantitative predictions of SU(3) symmetry breaking for $b\to s$ process in $bc$ sector.}
		\label{Tab:SU3_breaking_bc}%
		\begin{tabular}{l|c|c|c}
			\hline
			channels  & $\Gamma/\text{~GeV}$ (LFQM)  & $\Gamma/\text{~GeV}$ (SU(3))  & $|{\rm LFQM}-{\rm SU(3)}|/{\rm SU}(3)$ \tabularnewline
			\hline
			$\Xi_{bc}^{+}\to\Xi_{c}^{\prime+}e^{+}e^{-}$  & $4.54\times10^{-19}$  & $4.54\times10^{-19}$  & - -\tabularnewline
			$\Xi_{bc}^{0}\to\Xi_{c}^{\prime0}e^{+}e^{-}$  & $4.53\times10^{-19}$  & $4.54\times10^{-19}$  & $0\%$\tabularnewline
			$\Omega_{bc}^{0}\to\Omega_{c}^{0}e^{+}e^{-}$  & $7.42\times10^{-19}$  & $9.08\times10^{-19}$  & $18\%$ \tabularnewline
			\hline
			$\Xi_{bc}^{+}\to\Xi_{c}^{\prime+}\mu^{+}\mu^{-}$  & $3.97\times10^{-19}$  & $3.97\times10^{-19}$  & - - \tabularnewline
			$\Xi_{bc}^{0}\to\Xi_{c}^{\prime0}\mu^{+}\mu^{-}$  & $3.95\times10^{-19}$  & $3.97\times10^{-19}$  & $1\%$ \tabularnewline
			$\Omega_{bc}^{0}\to\Omega_{c}^{0}\mu^{+}\mu^{-}$  & $6.41\times10^{-19}$  & $7.94\times10^{-19}$  & $19\%$ \tabularnewline
			\hline
			$\Xi_{bc}^{+}\to\Xi_{c}^{\prime+}\tau^{+}\tau^{-}$  & $6.50\times10^{-20}$  & $6.50\times10^{-20}$  & - -\tabularnewline
			$\Xi_{bc}^{0}\to\Xi_{c}^{\prime0}\tau^{+}\tau^{-}$  & $6.45\times10^{-20}$  & $6.50\times10^{-20}$  & $1\%$ \tabularnewline
			$\Omega_{bc}^{0}\to\Omega_{c}^{0}\tau^{+}\tau^{-}$  & $9.12\times10^{-20}$  & $1.30\times10^{-19}$  & $30\%$ \tabularnewline
			\hline
		\end{tabular}
	\end{table}
	
	\begin{table}
		\caption{Quantitative predictions of SU(3) symmetry breaking for $b\to s$ process in $bc^{\prime}$
			sector.}
		\label{Tab:SU3_breaking_bcp}%
		\begin{tabular}{l|c|c|c}
			\hline
			channels  & $\Gamma/\text{~GeV}$ (LFQM)  & $\Gamma/\text{~GeV}$ (SU(3))  & $|{\rm LFQM}-{\rm SU(3)}|/{\rm SU}(3)$ \tabularnewline
			\hline
			$\Xi_{bc}^{\prime+}\to\Xi_{c}^{\prime+}e^{+}e^{-}$  & $1.27\times10^{-19}$  & $1.27\times10^{-19}$  & - -\tabularnewline
			$\Xi_{bc}^{\prime0}\to\Xi_{c}^{\prime0}e^{+}e^{-}$  & $1.26\times10^{-19}$  & $1.27\times10^{-19}$  & $1\%$\tabularnewline
			$\Omega_{bc}^{\prime0}\to\Omega_{c}^{0}e^{+}e^{-}$  & $2.11\times10^{-19}$  & $2.54\times10^{-19}$  & $17\%$ \tabularnewline
			\hline
			$\Xi_{bc}^{\prime+}\to\Xi_{c}^{\prime+}\mu^{+}\mu^{-}$  & $1.21\times10^{-19}$  & $1.21\times10^{-19}$  & - - \tabularnewline
			$\Xi_{bc}^{\prime0}\to\Xi_{c}^{\prime0}\mu^{+}\mu^{-}$  & $1.20\times10^{-19}$  & $1.21\times10^{-19}$  & $1\%$ \tabularnewline
			$\Omega_{bc}^{\prime0}\to\Omega_{c}^{0}\mu^{+}\mu^{-}$  & $2.01\times10^{-19}$  & $2.42\times10^{-19}$  & $17\%$ \tabularnewline
			\hline
			$\Xi_{bc}^{\prime+}\to\Xi_{c}^{\prime+}\tau^{+}\tau^{-}$  & $2.03\times10^{-20}$  & $2.03\times10^{-20}$  & - -\tabularnewline
			$\Xi_{bc}^{\prime0}\to\Xi_{c}^{\prime0}\tau^{+}\tau^{-}$  & $2.01\times10^{-20}$  & $2.03\times10^{-20}$  & $1\%$ \tabularnewline
			$\Omega_{bc}^{\prime0}\to\Omega_{c}^{0}\tau^{+}\tau^{-}$  & $2.91\times10^{-20}$  & $4.06\times10^{-20}$  & $28\%$ \tabularnewline
			\hline
		\end{tabular}
	\end{table}
	
	\subsection{Uncertainties}
	
	Also taking the process of $\Xi_{bb}^{0}\to\Xi_{b}^{0}$ as an example,
	the uncertainties caused by the model parameters and the single pole assumption will be given in
	this subsection. The error estimates for the form factors can be found
	in Table~\ref{Tab:error_ff}, in which the errors come from $\beta_{i}$,
	$\beta_{f}$ and $m_{di}$, respectively. The error estimates for the decay widths are listed below:
\begin{eqnarray}
\Gamma(\Xi_{bb}^{0}\to\Xi_{b}^{0}e^{+}e^{-}) & = & (1.98\pm0.49\pm1.21\pm0.13\pm0.26)\times10^{-19}\ {\rm GeV},\nonumber \\
\Gamma(\Xi_{bb}^{0}\to\Xi_{b}^{0}\mu^{+}\mu^{-}) & = & (1.92\pm0.48\pm1.18\pm0.14\pm0.26)\times10^{-19}\ {\rm GeV},\nonumber \\
\Gamma(\Xi_{bb}^{0}\to\Xi_{b}^{0}\tau^{+}\tau^{-}) & = & (3.72\pm0.96\pm2.52\pm0.51\pm1.28)\times10^{-20}\ {\rm GeV},\label{eq:error_width}
\end{eqnarray}
	where these errors come from $\beta_{i}$, $\beta_{f}$, $m_{di}$
	and $m_{{\rm pole}}$, respectively. The first three model parameters are all varied
	by 10\%, while $m_{{\rm pole}}$, which is responsible for the single pole assumption, is varied by 5\%. It can be seen from Table~\ref{Tab:error_ff} and Eqs.~(\ref{eq:error_width})
	that, the uncertainties caused by these parameters may be sizable.
	
	\begin{table}
		\caption{Error estimates for the form factors, taking $\Xi_{bb}^{0}\to\Xi_{b}^{0}$
			as an example. The first number is the central value, and the following 3 errors come from $\beta_{i}=\beta_{\Xi_{bb}^{0}}$, $\beta_{f}=\beta_{\Xi_{b}^{0}}$
			and $m_{di}=m_{(bu)}$, respectively. These parameters are all varied by 10\%. }
		\label{Tab:error_ff} %
		\begin{tabular}{c|c||c|c}
			\hline
			$F$  & $F(0)$  & $F$  & $F(0)$ \tabularnewline
			\hline
			$f_{1,S}^{\Xi_{bb}\to\Xi_{b}}$  & $0.141\pm0.018\pm0.036\pm0.002$  & $f_{1,A}^{\Xi_{bb}\to\Xi_{b}}$  & $0.138\pm0.018\pm0.035\pm0.002$ \tabularnewline
			$f_{2,S}^{\Xi_{bb}\to\Xi_{b}}$  & $-0.189\pm0.039\pm0.037\pm0.014$  & $f_{2,A}^{\Xi_{bb}\to\Xi_{b}}$  & $0.132\pm0.015\pm0.027\pm0.029$ \tabularnewline
			$f_{3,S}^{\Xi_{bb}\to\Xi_{b}}$  & $0.016\pm0.009\pm0.013\pm0.019$  & $f_{3,A}^{\Xi_{bb}\to\Xi_{b}}$  & $-0.068\pm0.006\pm0.007\pm0.022$ \tabularnewline
			\hline
			$g_{1,S}^{\Xi_{bb}\to\Xi_{b}}$  & $0.122\pm0.020\pm0.025\pm0.007$  & $g_{1,A}^{\Xi_{bb}\to\Xi_{b}}$  & $-0.030\pm0.004\pm0.007\pm0.001$ \tabularnewline
			$g_{2,S}^{\Xi_{bb}\to\Xi_{b}}$  & $0.056\pm0.016\pm0.045\pm0.030$  & $g_{2,A}^{\Xi_{bb}\to\Xi_{b}}$  & $-0.055\pm0.004\pm0.017\pm0.006$ \tabularnewline
			$g_{3,S}^{\Xi_{bb}\to\Xi_{b}}$  & $-0.406\pm0.088\pm0.225\pm0.120$  & $g_{3,A}^{\Xi_{bb}\to\Xi_{b}}$  & $0.261\pm0.019\pm0.078\pm0.022$ \tabularnewline
			\hline
			$f_{2,S}^{T,\Xi_{bb}\to\Xi_{b}}$  & $0.108\pm0.016\pm0.023\pm0.020$  & $f_{2,A}^{T,\Xi_{bb}\to\Xi_{b}}$  & $-0.066\pm0.013\pm0.013\pm0.010$ \tabularnewline
			$f_{3,S}^{T,\Xi_{bb}\to\Xi_{b}}$  & $0.091\pm0.018\pm0.018\pm0.013$  & $f_{3,A}^{T,\Xi_{bb}\to\Xi_{b}}$  & $0.134\pm0.024\pm0.026\pm0.011$ \tabularnewline
			\hline
			$g_{2,S}^{T,\Xi_{bb}\to\Xi_{b}}$  & $0.128\pm0.012\pm0.036\pm0.002$  & $g_{2,A}^{T,\Xi_{bb}\to\Xi_{b}}$  & $-0.049\pm0.005\pm0.012\pm0.001$ \tabularnewline
			$g_{3,S}^{T,\Xi_{bb}\to\Xi_{b}}$  & $0.156\pm0.122\pm0.020\pm0.012$  & $g_{3,A}^{T,\Xi_{bb}\to\Xi_{b}}$  & $0.032\pm0.010\pm0.012\pm0.002$ \tabularnewline
			\hline
		\end{tabular}
	\end{table}
	
	\section{Conclusions}
	
	In our previous work, we have investigated the weak decays of doubly
	heavy baryons for 1/2 to 1/2 case and for 1/2 to 3/2 case.
	As a continuation, we investigate the FCNC processes in this work.
	Light-front approach under the diquark picture is once again adopted
	to extract the form factors. The same method was applied to study the singly heavy baryon decays and reasonable results
	were obtained~\cite{Zhao:2018zcb}. The extracted form factors are
	then applied to study some observables in these FCNC processes. We find that most of the
	branching ratios for $b\to s$ processes are $10^{-8}\sim10^{-7}$,
	while those for $b\to d$ processes are $10^{-9}\sim10^{-8}$, which
	are roughly one order of magnitude smaller than those in mesonic sector. This is because we believe that the lifetime of the doubly heavy baryon is roughly one order of magnitude smaller than that of $B$ meson. SU(3) symmetry and sources of symmetry breaking are discussed. The error estimates are also investigated.
	
	\section*{Acknowledgements}
	
	The authors are grateful to Prof.~Wei Wang for valuable discussions
	and constant encouragements. This work is supported in part by National
	Natural Science Foundation of China under Grant Nos.~11575110, 11655002,
	11735010, Natural Science Foundation of Shanghai under Grant No.~15DZ2272100.

\end{document}